# Recommended Implementation of Quantitative Susceptibility Mapping for Clinical Research in The Brain: A Consensus of the ISMRM Electro-Magnetic Tissue Properties Study Group

**QSM Consensus Organization Committee (in alphabetical order)**, Berkin Bilgic[1], Mauro Costagli[2,3], Kwok-Shing Chan[4], Jeff Duyn[5], Christian Langkammer[6], Jongho Lee[7], Xu Li[8,9], Chunlei Liu[10,11], José P. Marques[4], Carlos Milovic[12], Simon Robinson[13], Ferdinand Schweser[14,15,*], Karin Shmueli[16], Pascal Spincemaille[17], Sina Straub[18], Peter van Zijl[8,9], Yi Wang[19,*], **ISMRM Electro-Magnetic Tissue Properties Study Group****

* Corresponding authors

## Author Affiliations

1 – Martinos Center for Biomedical Imaging, Massachusetts General Hospital and Harvard Medical School, Charlestown, MA, United States
2 – Department of Neuroscience, Rehabilitation, Ophthalmology, Genetics, Maternal and Child Sciences (DINOGMI), University of Genoa, Genoa, Italy
3 – Laboratory of Medical Physics and Magnetic Resonance, IRCCS Stella Maris, Pisa, Italy
4 – Donders Institute for Brain, Cognition and Behaviour, Radboud University, Nijmegen, the Netherlands
5 – Advanced MRI Section, NINDS, National Institutes of Health, Bethesda, MD, United States
6 – Department of Neurology, Medical University of Graz, Graz, Austria
7 – Department of Electrical and Computer Engineering, Seoul National University, Seoul, Republic of Korea
8 – Russell H. Morgan Department of Radiology and Radiological Science, Johns Hopkins University School of Medicine, Baltimore, MD, United States
9 – F.M. Kirby Research Center for Functional Brain Imaging, Kennedy Krieger Institute, Baltimore, MD, United States
10 – Department of Electrical Engineering and Computer Sciences, University of California, Berkeley, CA, USA
11 – Helen Wills Neuroscience Institute, University of California, Berkeley, CA, USA

12 – School of Electrical Engineering (EIE), Pontificia Universidad Catolica de Valparaiso, Valparaiso, Chile
13 – High Field MR Centre, Department of Biomedical Imaging and Image-Guided Therapy, Medical University of Vienna, Austria
14 – Buffalo Neuroimaging Analysis Center, Department of Neurology, Jacobs School of Medicine and Biomedical Sciences at the University at Buffalo, Buffalo, NY, USA
15 – Center for Biomedical Imaging, Clinical and Translational Science Institute at the University at Buffalo, Buffalo, NY, United States
16 – Department of Medical Physics and Biomedical Engineering, University College London, London, UK
17 – MRI Research Institute, Department of Radiology, Weill Cornell Medicine, New York, NY, United States
18 – Department of Radiology, Mayo Clinic, Jacksonville, FL, United States
19 – MRI Research Institute, Departments of Radiology and Biomedical Engineering, Cornell University, New York, NY, United States

# Abstract

This article provides recommendations for implementing quantitative susceptibility mapping (QSM) for clinical brain research. It is a consensus of the ISMRM Electro-Magnetic Tissue Properties Study Group. While QSM technical development continues to advance rapidly, the current QSM methods have been demonstrated to be repeatable and reproducible for generating quantitative tissue magnetic susceptibility maps in the brain. However, the many QSM approaches available give rise to the need in the neuroimaging community for guidelines on implementation. This article describes relevant considerations and provides specific implementation recommendations for all steps in QSM data acquisition, processing, analysis, and presentation in scientific publications. We recommend that data be acquired using a monopolar 3D multi-echo GRE sequence, that phase images be saved and exported in DICOM format and unwrapped using an exact unwrapping approach. Multi-echo images should be combined before background removal, and a brain mask created using a brain extraction tool with the incorporation of phase-quality-based masking. Background fields should be removed within the brain mask using a technique based on SHARP or PDF, and the optimization approach to dipole inversion should be employed with a sparsity-based regularization. Susceptibility values should be measured relative to a specified reference, including the common reference region of whole brain as a region of interest in the analysis, and QSM results should be reported with – as a minimum – the acquisition and processing specifications listed in the last section of the article. These recommendations should facilitate clinical QSM research and lead to increased harmonization in data acquisition, analysis, and reporting.

# Keywords

# 1. Introduction

Brain quantitative susceptibility mapping (QSM) has been increasingly performed to identify calcifications and study iron, myelin, and oxygen consumption changes associated with normal brain development or aging and with neurological disease. Diseases of interest include hemorrhagic stroke, multiple sclerosis, Alzheimer's disease, Parkinson's disease, amyotrophic lateral sclerosis, and tumors [1–15]. QSM is also increasingly being used in psychiatric disorders such as psychosis [16,17] or depression [18,19], where it may reflect neurotransmitter or metabolic imbalances. Changes in brain susceptibility have also been associated with alcohol consumption [20–22], potentially informing the neural mechanisms through which alcohol affects the brain. Further, QSM has been used for the differentiation between hemorrhages and calcifications [23–26], anatomical visualization [27–29] and improved segmentation [30,31], and for presurgical mapping in deep brain stimulation because it depicts deep gray nuclei targets with exquisite contrast and superior iron source sharpness as compared to other approaches such as $T_2$ and $T_2^*$-weighted imaging [27,32–36]. In conjunction with $R_2$ or $R_2^*$ mapping, QSM may also be used to separate diamagnetic myelin and calcification from paramagnetic iron within a voxel [37–46].

While QSM techniques and their biomedical applications have been extensively reviewed [3,6,7,47–56,57(p31),58,59], and the QSM research community has held two challenges attempting to identify the best susceptibility calculation algorithm [60–63], there has been no community consensus or white paper on how to best perform brain QSM in the clinical setting. Some vendors have started implementing research-level QSM pipelines for their systems, nevertheless QSM is not yet a product on most magnetic resonance imaging (MRI) systems. On the other hand, demands for a robust "end-to-end" QSM recipe covering acquisition and processing through to presentation in scientific publications have been growing strongly from the neurological imaging community. For example, the need for a QSM consensus method was highlighted at the group discussion of the February 2022 workshop of the North American Imaging in Multiple Sclerosis (NAIMS) Cooperative, as QSM is particularly useful for depicting the paramagnetic rim of chronic active multiple sclerosis lesions [64–72]. In response to this need, QSM investigators in the Electro-Magnetic Tissue Properties Study Group (EMTP SG) of the International Society for Magnetic Resonance in Medicine (ISMRM) established a QSM Consensus Organization Committee to define the scope of the recommendations and determine the current consensus.

This present article presents the scope of the recommendations, describes the approaches used to obtain consensus, reviews of each step of the QSM reconstruction process, and summarizes its specific consensus recommendations. The technical sections present, first, a general overview of the subject matter in the context of the QSM reconstruction process and then the consensus recommendation. Readers interested only in the consensus recommendations may skip the overview sections and proceed directly to the respective recommendation subsections. Readers interested in the technical details and justification of the recommendations are encouraged to read the overview sections. All consensus recommendations are systematically enumerated to facilitate referencing of the individual recommendations in the literature. Where needed, a sub-section with additional considerations is presented. The paper closes with a summary of the recommendations.

## 1.1 Scope

This consensus paper was written as a guide on how to implement QSM in clinical research for clinical researchers not familiar with the technical nuances of QSM. The implementation of the recommendations requires the assistance of technical personnel such as medical physicists or vendors with in-depth knowledge of the MRI scanner and with expertise in image processing and analysis. The main purpose of this paper is to increase the use of QSM in clinical trials and in different patient groups. Guidelines for clinical practice are beyond the scope of the current paper and would require further study.

Recommendations represent a consensus among QSM experts and the community of experienced QSM users at the time of publication. For clarity, the committee decided that the emphasis of the paper would be on 3 tesla (T), the field strength most widely used in clinical brain research, and that general guidance would be provided for other field strengths (1.5 T and 7 T) where possible. Due to the focus on clinical applications, including clinical trials, the recommendations prioritize robustness and simplicity over state-of-the-art acquisitions and processing algorithms, and provide a starting point for application-specific improvements. Areas in which the committee could not yet reach a consensus (e.g., due to insufficient evidence) are indicated as such in the paper.

## 1.2 Organization

The article is organized into eight sections, covering data acquisition, image processing, image analysis, and presentation of QSM studies in scientific publications (Fig.1). Data acquisition is split into two sections: 1) pulse sequences and protocol, and 2) coil combination, saving, and exporting; image processing is split into four sections corresponding to four processing steps: phase unwrapping and echo combination, creation of masks, background field removal, and dipole inversion; the image analysis section focuses on the analysis of susceptibility maps; and the last section on presentation covers reporting in scientific publications. The contributions of the members of the QSM Consensus Organization Committee are listed in the Supplementary Materials I, Section S1.1 along with an overview of the history and approach of the initiative (Section S1.3). The Acknowledgement section lists all individuals who contributed significantly to the consensus recommendations in verbal or written form.

The paper is accompanied by harmonized pulse sequence protocols for several major platforms (current software versions in 2022) as well as sample code to perform the recommended processing steps (see Supplementary Materials II) using the Sepia toolbox [73] [available at https://github.com/kschan0214/sepia]. In addition, curated online resources are provided and will be updated as the field progresses. This information is detailed in the Supplementary Materials.

## 1.3 Background

Tissue magnetic susceptibility is directly related to tissue chemical composition. A substance gaining a magnetization opposing an externally applied magnetic field, e.g. calcium, is called diamagnetic. Diamagnetism is due to electron orbit perturbation. A substance gaining a magnetization in the same direction as the external field, e.g. iron, is called paramagnetic. Paramagnetism is due to spin alignment of unpaired electrons, which have a magnetic moment 658 times greater than that of the hydrogen nucleus. Most biological tissues, like the human brain, have a susceptibility close to that of water, which is diamagnetic with a value of roughly -9 parts per million (ppm). In the brain, iron deposition makes tissue less diamagnetic than water, and calcium deposition and the presence of myelin make tissue more diamagnetic than water [4,11,15,74–76].

The static magnetic field of magnetic resonance imaging (MRI) scanners, $B_0$, affects electrons in tissue and, in linear materials, induces a magnetization proportional to $B_0$ with the proportionality constant defined as magnetic susceptibility $\chi$ of the tissue. The induced tissue magnetization generates a field that is nonlocal and extends into the space surrounding the magnetization source, thereby inducing field inhomogeneities. These inhomogeneities can be described mathematically as a convolution of the susceptibility distribution with the unit dipole kernel [77–79]. During MRI signal generation, water proton spins experience this tissue field. In particular, gradient-recalled echo (GRE) imaging is very sensitive to this tissue field. Consequently, there is complex destructive interfering or dephasing by the tissue field variation within a voxel seen as hypointensities or blooming susceptibility artifacts on the GRE magnitude images, which is known as $T_2^*$-weighted or susceptibility-weighted imaging [80–82]. The GRE phase images, traditionally discarded, represent the tissue field inhomogeneities averaged over the voxel. The process by which tissue magnetic susceptibility is computed from the magnetic field measurement is called quantitative susceptibility mapping (QSM) [83,84]. Several detailed review papers have been published about the technical foundations and clinical applications of QSM, and we refer the interested reader to these articles for a comprehensive overview of the technique [5,13,49,50,52–56,59,85–93].

## 2. Pulse Sequences and Protocol

This section describes recommendations to robustly acquire data for QSM.

### 2.1. Overview

Most of the major MRI manufacturers, if not all of them, provide a radiofrequency (RF)-spoiled 3D multi-echo GRE pulse sequence from which one can obtain phase images for QSM in addition to the standard $T_2^*$-weighted magnitude images. The GRE sequence is probably the most elementary sequence in the MRI sequence tree, consisting in its simplest form of an RF excitation pulse followed by acquisition of a gradient recalled echo, which for the multi-echo variant is repeated at various echo times (TEs) before the next excitation pulse is applied. This sequence is often used for calibration steps, such as obtaining a field map for $B_0$ shimming.

Optimizing a GRE sequence to maximize QSM contrast shares many of the principles for optimizing $T_2^*$-weighted contrast-to-noise ratio or $T_2^*$ mapping [94]. We recommend the following principles for designing a QSM acquisition protocol:

- Aim to set the last TE equal to at least the $T_2^*$ value of the tissue of interest. For example, if targeting deep gray matter, the $T_2^*$ of the putamen can be a good reference (55, 30 and 16 ms at 1.5, 3 and 7 T, respectively [95]). The first echo time (TE1) should be as short as possible and the spacing between consecutive echoes (ΔTE) should be uniform throughout the echo train.
- Use the minimum repetition time (TR) and set the flip angle to the Ernst angle ($\theta_{Ernst} = \cos^{-1}(e^{-TR/T_1})$) for the target region. Most deep gray matter structures have a $T_1$ close to that of white matter, and white matter represents the largest volume fraction in the brain. Hence, the Ernst angle may be calculated for white matter or deep gray matter with $T_1$ = 650, 850 and 1220 ms at 1.5, 3 and 7 T respectively [96].
- Use three or more monopolar echoes. While QSM can be achieved with one echo and two is the minimum number of echoes needed to separate the intrinsic transmit RF phase from the magnetic field-induced phase (see the Section 3 below), the use of a larger number of echoes (e.g., 5 echoes) will benefit the phase signal-to-noise ratio (SNR) for a range of tissues [97]. As phase SNR is maximal when TE = $T_2^*$, the use of multiple echoes ensures that high SNR field estimates are obtained for both short apparent $T_2^*$ (venous blood or tissues close to air-tissue interfaces) and other tissues with longer $T_2^*$ values.
- Use the minimum readout bandwidth (BW) that generates acceptable distortions. At 3T, 220 Hz/pixel is often sufficient (two-pixel fat-water shift). Such acquisitions negate the need to use fat suppression for brain applications. Note that as the BW is increased the ratio of the readout duration (1/BW) and echo spacing decreases (as a consequence of the required time to rewind the gradients), which results in a reduction in the SNR of the acquisition. The best trade-off depends on the system's gradient amplitude and slew rate.
- Use isotropic voxels of at most 1 mm to avoid susceptibility underestimation reported to occur at larger voxel sizes [98].
- Use 3D acquisition instead of 2D acquisition to avoid potential slice-to-slice phase discontinuities in 2D phase maps that mainly occur in interleaved acquisitions and require additional processing or to avoid slice crosstalk in non-interleaved acquisitions [99]. Note that the 3D spatial coherence of the obtained phase/field maps is of great importance as QSM is ultimately a 3D spatial deconvolution process.
- Use a monopolar gradient readout (fly-back) to avoid geometric mismatch and eddy current related phase problems between even and odd echoes in bipolar acquisitions since these require additional correction in the pipeline used [100].

- Consider using flow compensation when targeting vessels (e.g., oxygenation studies), but note that flow compensation is often only available and guaranteed for the first echo, while flow artifacts increase in later echoes. Flow compensation also reduces the number of echoes achievable as it increases the minimum TE, assuming a fixed BW. In comparison, flow compensation has a much smaller effect on the resulting QSM values than the choice of QSM processing pipeline [101].

The following points focus on practical aspects of image orientation, FOV, and acceleration to achieve a whole brain data acquisition in approximately 6 minutes for clinical research on a 3 T system with as few as 8 receive channels using a standard 3D multi-echo RF-spoiled GRE sequence:

- Patients should be scanned in the supine position with the head straight to minimize variability in white matter susceptibility related to myelin magnetic susceptibility anisotropy [102,103] and microstructural water compartmentation [104].
- For whole brain acquisition (including cerebellum), it is recommended to use isotropic voxels [98] and prescribe the imaging volume with the readout direction along the anterior commissure - posterior commissure (AC-PC) line (oblique axial orientation), which reduces the number of phase encoding steps and, consequently, the total scan time [105]. Setting the readout in this direction also restricts eye movement artifacts in the left-right direction, outside of the brain. Alternatively, sagittal acquisitions (readout along the head-foot direction) can be beneficial when the region of interest includes the brain stem but will require larger acceleration factors. The use of tilted imaging slab orientations (oblique axial or oblique sagittal) requires that the data header information (see Section 3 below) specifying the exact slab orientation is supplied to the subsequent processing pipeline, otherwise artifacts will occur in the resulting susceptibility map [106].
- Parallel imaging and partial k-space coverage can be used to reduce acquisition time, as for any structural MRI acquisition. Below we give general guidelines for GRE brain imaging:
    - If accelerating in only one direction, use acceleration factors <3. If available, non-integer acceleration factors allow fine-tuning the acceleration factors to the available equipment.

- If available, elliptical k-space sampling (elliptical k-space shutter) and/or partial k-space acquisition (e.g., 7/8) with zero-filling can be used to reduce scan time but note that this can result in a reduction of the effective spatial resolution.
- If an RF coil with 32 or more channels is available, acceleration in both phase encoding directions by a factor of 2 (2x2) can be used, preferably with CAIPI patterns [107].
- If available, similar or higher acceleration factors compared to conventional accelerated imaging may be obtained with incoherent sampling using compressed sensing reconstruction methods [108] or with deep learning reconstruction methods [109].

## 2.2. Consensus Recommendation

The recommendation in this section is based on protocols currently available on commercial scanners that do not require a special research agreement with the scanner manufacturer:

2.a. 3D multi-echo RF-spoiled GRE with monopolar readout. Three or more echoes should be acquired and the TE range should include the $T_2^*$ times of the target tissues.
2.b. Minimum TR (given selected TEs).
2.c. Flip angle should be set to the Ernst angle for target tissues (e.g., white matter).
2.d. Whole brain coverage.
2.e. Resolution should be isotropic with a voxel edge length of at most 1 mm non-interpolated at 3T.
2.f. Use accelerated imaging methods (e.g., parallel imaging).
2.g. Use coil arrays with a large number of elements covering the whole brain.

The Supplementary Materials I, Section S1.5 contain sample protocols for 1.5T, 3T, and 7T.

## 2.3. Additional Considerations

Acquisitions for QSM reconstructions can be performed at all clinical field strengths with minimal image distortions using the protocol recommendations provided here, although higher field strengths provide several benefits. Susceptibility contrast and contrast-to-noise ratio (CNR) increase with the static magnetic field once relaxation time changes are considered [110], and acquisition can be more time efficient at higher fields due to the shorter $T_2^*$. Our recommended protocols implicitly integrate a compensation for lower SNR at lower $B_0$ through lower spatial resolution, while keeping the total acquisition times similar. While useful data can be obtained at all field strengths, finer anatomical details in certain applications, such as brain lesions in multiple sclerosis [111], might be difficult to visualize at 1.5T.

Advanced sequences such as segmented 3D-EPI [112,113] or Wave-CAIPI [114] allow drastic decreases in acquisition time but are not available as commercial sequences on all vendors and may require a research agreement with the scanner manufacturer. Nevertheless, it should be noted that some of these methods may have constraints on the number of echoes that can be acquired and 3D-EPI sequences will have increased spatial distortions that may have to be corrected for and that may result in additional artifacts that subsequent processing steps such as background field removal need to account for.

# 3. Coil Combination, Saving, and Exporting

This section recommends effective and practical solutions for generating phase images that can be used for QSM and provides guidance on how to save and convert the format of the phase images in preparation for QSM analysis.

## 3.1. Overview

Modern MRI systems use phased array coils made up typically of 12, 32 or 64 elements, which provide higher SNR than a single birdcage coil and facilitate parallel imaging [115–117]. However, the images from individual coil elements are sensitive to only a part of the FOV and need to be combined to generate a single image – a process that requires consideration of the differences between the coil signals.

Neglecting phase wraps (see Section 4), the phase measured with a particular RF coil element $c$ in a GRE sequence, $\varphi^c$, is given by: $\varphi^c = \varphi_B(\text{TE}) + \varphi_0^c$. The first part, $\varphi_B(\text{TE})$, is the phase shift caused by the deviation $\Delta B_0$ of the magnetic field from the uniform main magnetic field, $B_0$. Neglecting non-linear effects (discussed in Section 9.1.6), $\varphi_B(\text{TE})$ evolves linearly over time and is the term that is relevant for QSM. The second part, $\varphi_0^c$, is the phase at TE=0, known as the phase offset (or initial phase) of coil element $c$. This contribution comprises effects that are common to all the RF receive coils in a phased array, such as the phase of the transmit RF field $B_1^+$, the effects of tissue conductivity (Liu et al. 2017), gradient delays and eddy current effects, and contributions that are unique to each receive coil, such as the coil sensitivity. Coil-dependent phase offsets must be removed prior to a complex summation of the coil signals, as shown in Figure 2, to avoid destructive interference. Destructive interference leads to reduced SNR and unphysical phase wraps in regions of the image, which cause artifacts in QSM [118]. Complete destructive interference is often associated with phase wraps referred to as "open-ended fringe lines" or "phase singularities" (see Figure 3, left). These ill-behaved wraps do not represent isophase contours and cannot be fully removed by unwrapping (see Section 3 below).

All major 3T MR system vendors have effective solutions for generating phase images from RF array coils on their current software platforms. Most of these techniques remove individual coil sensitivities, which are estimated by referencing to a coil with a relatively homogenous sensitivity over the object (usually the body coil). The reference data is acquired in a separate, fast, automated measurement, and the coil sensitivity correction is carried out on complex data either in k-space [115] or image space [119] before extraction of the phase. To combine the signals from each coil element, we recommend using the available 'on-console' vendor solutions listed at the end of this section. These methods may not be available on systems running older software versions, in which case phase images can be reconstructed from 'raw' (k-space) data offline or, alternatively, phase images can be saved and exported separately for each coil element (e.g., as DICOM-format image data) to allow an appropriate coil combination method to be applied offline. These offline solutions may require additional efforts in 1) dealing with the export and transfer of very large files or a large number of files, which can be problematic in a clinical setting and 2) obtaining research agreements with vendors for proprietary information on the raw data format and on special routines for performing the image reconstruction offline. Several offline possibilities are nonetheless described in the Supplementary Materials III.

A survey of the members of the committee, conducted as part of this study, showed that most respondents use the coil combination methods listed in Section 3.2. A small proportion of users of Siemens systems, particularly with those running older software or those imaging at 7T, use ASPIRE [120] or the separate channel coil export with an offline solution. In the broader community, many groups continue to use offline solutions for consistency with 7T (where no body coil-based solution exists) or to provide consistency with older studies.

For all systems, data should be exported in DICOM format. However, most QSM tools require data to be in NIfTI format, and a number of tools are available to perform the necessary conversion. We recommend DCM2NIIX (https://github.com/rordenlab/dcm2niix), which is a well-maintained open-source software with compiled versions for macOS, Linux and Windows. It has been tested on data from a range of scanners [121], including the example data provided with this project. The following alternative programs fulfill a similar function: DICM2NII (https://github.com/xiangruili/dicm2nii), MRIConvert (https://lcni.uoregon.edu/downloads/mriconvert/mriconvert-and-mcverter), and Dicomifier (https://github.com/lamyj/dicomifier). In our survey, DCM2NIIX was the most popular conversion tool, but with a significant spread across the others listed. Format conversion with DCM2NIIX and the analysis steps recommended in this paper preserve the correct image orientation. It should be noted, though, that for non-vendor imaging sequences, some conversion tools and non-standard processing steps (or some combination of these) may not save or handle header parameters correctly. If any of these are used, the researcher should check that left-right flips have not been introduced into images (see Section 8.1.1).

If manufacturers' product susceptibility weighted imaging (SWI) sequences [122–124] are used for QSM acquisition, it should be noted that many of these do not allow direct saving or export of unfiltered phase images. Some do allow saving of the processed (e.g., homodyne filtered) phase data, but these are unsuitable for QSM. If raw k-space data are available from SWI scans, it may be possible to generate phase images usable for QSM from these (see Supplementary Materials III). Where available, we recommend that clinical researchers use scanners and sequences which allow saving and exporting of the original unprocessed phase wherever possible, as this greatly facilitates (retrospective) clinical QSM studies and accelerates clinical translation of QSM.

It is important to make sure that phase images are scaled correctly and converted to the correct data type for subsequent analysis. The analysis pipeline supplied with this paper works with a range of data types and arbitrary phase scaling, but some phase unwrapping algorithms and the nonlinear complex fitting algorithm described in Section 4 require phase to be saved as floating-point numbers scaled between $-\pi$ and $\pi$, and this rescaling and data type conversion needs to be performed by the user. Further, note that the sign convention is different for phase values on different vendor systems [125] and this needs to be checked and corrected where necessary so that relatively paramagnetic tissues (e.g., iron-rich deep-brain regions) have positive susceptibility values in the resulting susceptibility maps.

## 3.2. Consensus Recommendations

We provide recommendations for combining phase images from array coils and saving phase data for each of the major manufacturers, listing the software versions for which these solutions have been tested and whether a research agreement is required. Detailed step-by-step descriptions and solutions for older systems are provided in Supplementary Materials III.

3.a The recommended solution for saving phase images are, for
- Canon: SPEEDER, a version of SENSE which is available from MPower version 2.3 onwards and allows phase images to be reconstructed through a vendor-provided service password.
- GE: ASSET, a SENSE-similar solution which reconstructs magnitude, phase, real, and imaginary images without a research key on platforms MR30 onward.
- Philips: SENSE, which provides well-combined phase images [116] without the need for a research key from software version 5 onwards.
- Siemens: "Adaptive-combined with prescan normalize" [126], which is available from software version VE11 onward in the product GRE sequence.
- United Imaging: an inter-coil referencing and weighted correction approach which is available from software version v9 without the need for a research key.

3.b Exporting data: Data should be exported in (classic) DICOM format.

3.c Format conversion: if the analysis pipeline requires NIfTI data, DICOM data should be converted to NIfTI using DCM2NIIX.

## 3.3. Additional Considerations

# 4. Phase Unwrapping and Echo Combination

This section describes the methods used to resolve phase aliasing and calculate a field map from multi-echo GRE data.

## 4.1. Overview

MRI phase measurements are constrained to an interval of $2\pi$ and are, therefore, subject to phase wraps or phase aliasing artifacts, i.e., the measured phase $\varphi = (\varphi_B(\text{TE}) + \varphi_0) \mod 2\pi$. Such phase wraps introduce a phase difference of an integer multiple of $2\pi$ between the measured phase, $\varphi$, and the true phase $\varphi_B(\text{TE}) + \varphi_0$. Phase wraps are usually visible as discontinuous phase jumps in the phase images (Figs. 3 and 4). To a first-order approximation, $\varphi_B(\text{TE}) = 2\pi\gamma\text{TE} \cdot \Delta B_0$, where $\gamma$ is the proton gyromagnetic ratio. To obtain an accurate estimate of the field shift, $\Delta B_0$, for QSM, both phase wraps and the phase offset $\varphi_0$ need to be removed from the measured phase, $\varphi$ [54]. The coil combination methods recommended in the previous section remove the coil-specific contributions to the phase offset, $\varphi_0$, but leave other non-$B_0$-related contributions in $\varphi_0$, common to all coils, which should be removed by the QSM processing pipeline. Note that the background field removal step (see Section 6 below) removes only harmonic fields (those which satisfy Laplace's equation) within the brain region and cannot completely remove $\varphi_0$ as it contains both harmonic and non-harmonic components [55](Schweser et al., 2017). Therefore, $\varphi_0$ must be explicitly removed for accurate QSM [127].

Over the years, different phase unwrapping methods have been adapted, refined and applied to MR phase imaging; including time-domain unwrapping methods (with multi-echo acquisition) such as CAMPUS [128] and UMPIRE [129], and spatial-domain unwrapping methods such as region-based PRELUDE [130], SEGUE [131], and SPUN [132], path-based best-path unwrapping [133] and ROMEO [134], and Laplacian unwrapping [135]. The Laplacian unwrapping method is robust and gives wrap-free phase results even with low SNR but can result in high-frequency errors that propagate into susceptibility maps that are hard to detect visually [136]. It is also noted that Laplacian unwrapping only gives an approximation of the underlying unwrapped phase, especially when using the commonly used Fourier-based implementation, while region-based and path-based unwrapping give quantitatively more accurate estimates of the unwrapped phase [54,134]. Region-based and path-based methods are termed "exact unwrapping methods" below as they give the exact value of the unwrapped phase [134]. When comparing exact unwrapping methods to Laplacian unwrapping, unwrapping errors (e.g., in veins and hemorrhages) are observed to be smaller in the former, improving QSM quantification accuracy, e.g., for oxygenation estimation [137,138]. Therefore, we recommend using exact unwrapping methods.

Multi-echo phase images, e.g., acquired using the recommended protocol (see Section 2), can be combined to achieve a more accurate estimate of the underlying field shift, $\Delta B_0$, than can be obtained from single-echo phase images [97]. This is because combining multi-echo phase images can remove the phase offset contribution and give higher SNR in the estimated tissue field and susceptibility maps (Fig. 4). The optimum approach may depend on the application, but two echo combination methods have been widely used for QSM – nonlinear complex data fitting [139] and weighted echo averaging [94].

The nonlinear complex data fitting approach takes into account the Gaussian noise in the complex images [139] and estimates the field shift, $\Delta B_0$, and phase offset, $\varphi_0$, together as parameters from fitting the complex MR signal over multiple echoes, with the requirement of having acquired three or more echoes [139]. This approach usually needs spatial phase unwrapping to be performed after the fitting, i.e., on a scaled field-shift estimate, e.g., $2\pi\gamma \cdot \delta TE \cdot \Delta B_0$, with $\delta TE$ being the echo spacing, wrapped again between $-\pi$ and $+\pi$, as it resolves phase wraps in the temporal dimension. Nonlinear complex data fitting is more robust than linear phase fitting against phase noise at long TEs and around large susceptibility sources, e.g., veins and hemorrhages.

If echoes are acquired over a useful range of TE (depending on $T_2^*$ values of the tissues of interest, see Section 2), the weighted echo averaging approach gives higher SNR for estimating the field shift, $\Delta B_0$, than the complex data fitting approach, which estimates multiple parameters [94,140,141]. Unlike nonlinear complex data fitting, this approach needs the phase data at each TE to be unwrapped first. It also requires explicit removal of the phase offset through subtraction of the estimated phase offset from the phase measured at each TE (for example as in MCPC-3D-S and ASPIRE, where the phase offset can be estimated by extrapolating the linear phase evolution to zero echo time) [120]. Unwrapping errors at longer TEs in voxels with large field shifts and more pronounced noise can be reduced by the "template" unwrapping approach used in ROMEO, which performs path-based spatial unwrapping on an early echo and unwraps other echoes on the basis of the expected linear phase evolution [134]. This, combined with weighted echo averaging, reduces the effect of such errors in the estimated field map.

To improve QSM quality, the spatial noise map generated from nonlinear complex fitting [139], or the phase "quality map" calculated in path-based unwrapping [134] can be used to mask out voxels with unreliable phase (see Section 5).

Most echo combination methods assume linear phase evolution over TE, ignoring non-linear phase evolution due to microstructure-related compartmentalization effects or biased sampling of the sub-voxel field perturbation (see Section 9.1.6), flow, or signal dropout [142]. Advanced modeling of the phase evolution over time may provide further information about tissue composition and microstructure [37,104,143–148], but is beyond the scope of this paper.

## 4.2. Consensus Recommendations

    4.a.        Use an exact phase unwrapping method.

    4.b.        Perform echo combination before background field removal.

    4.c.        The optimal pipeline for phase unwrapping and echo combination depends on the acquisition and application. We recommend using either nonlinear complex data fitting followed by spatial phase unwrapping, or weighted echo averaging after template phase unwrapping and explicit phase offset removal.

## 4.3. Additional Considerations

# 5. Creation of masks

This section provides recommendations on creating masks for background field removal (Section 6), dipole inversion (Section 7), and visualization (Section 9).

## 5.1. Overview

Masking is often overlooked when describing a QSM pipeline but is a crucial step [149], particularly for background field removal (see Section 6). Masking refers to selecting a region of interest (ROI) within the whole field of view and applying a process or function only within this ROI. In QSM, field maps, $\Delta B_0$, are masked primarily because most background field removal algorithms require a mask. In general, masks should cover the largest ROI possible to prevent exclusion of brain tissue with a sufficient signal-to-noise ratio to have reliable phase/field values. This is of special concern for studies of the cortex and the brainstem near the brain border or air-tissue interfaces. Unreliable field map data is composed mostly of extremely noisy voxels resulting from phase noise in regions with very low MRI signal or rapid signal decay. The noise distribution in phase images (and hence, field maps) is generally non-Gaussian and depends on the local magnitude of the signal [150]. In practice, regions with very low MRI signal yield phase noise uniformly distributed throughout the whole -π to π range, obscuring any underlying phase contrast information [150].

Masking is a binary operation. Voxels with mask values of 1 (or Boolean "true") are included in the selected ROI and voxels with mask values of zero (or Boolean "false") are excluded. Masks may be created by using heuristic thresholding operations on available subject images, including magnitude images, $T_2^*$ maps, quality maps, or SNR maps. Masks created from differently thresholded images may also be joined or combined to exclude or include regions based on different criteria. In addition, segmentation algorithms may be based on pre-learned shapes or on the optimization of functionals [151–157]. In particular, the Brain Extraction Tool (BET) [158] from the FMRIB Software Library (FSL) is a widely-used method for brain masking (skull stripping), although it may fail when pathologies or injuries are present [159,160]. BET is a magnitude-image based algorithm that effectively removes non-brain tissues, air, and bone from magnitude images of the head. When acquiring multiple echoes, using the last-echo magnitude image for BET masking is robust to remove regions with rapid signal dropout [161], which is undesirable if such regions are of interest. A more balanced approach with a larger ROI selection is achieved by using magnitude images combined across TEs (e.g., using sum of squares or weighted averaging). Alternatively, the magnitude image of the first echo can be used for brain extraction, with the use of a phase-quality map to further exclude voxels with unreliable phase values, as described below. Alternatives to BET include standard template-based brain-extraction [162]. Deep learning segmentation alternatives may also be considered, as this is a rapidly developing field [163–166].

QSM is also vulnerable to errors and artifacts arising from unreliable phase data that may not be directly reflected in the corresponding magnitude data. These may be caused by coil combination errors, flow in vessels, and other factors. For this reason, it has been proposed to use phase-based quality maps in addition to magnitude-based masking to refine masking [161]. A straightforward method to obtain a phase quality map is to threshold the inverse of the noise map provided by the complex nonlinear multi-echo fitting algorithm (described in Section 4) at its mean value [139,161]. The thresholding at the mean value maintains an adequate number of voxels, as it is applied to the entire field of view (FOV), and the distribution of values exhibits bimodal characteristics. This approach effectively distinguishes between reliable and unreliable voxels, serving as a suitable initial approximation. However, modifying the threshold factor (e.g., by multiplying it by 1.2) may yield further enhancement in the results. Some exact phase unwrapping algorithms provide an alternative source of phase-based quality maps [134,167]. Setting a threshold for phase-based quality maps can help to identify voxels within the brain with unreliable phase values, and to provide a better estimation of the brain boundary [149,161,168].

Most phase unwrapping and echo combination algorithms do not require masking [54] but suppressing extraneous data by masking can speed up some algorithms and improve their robustness. In contrast, almost all background field removal algorithms require masking to define the region of interest, outside which the susceptibility sources are classified as background field sources (see Section 6) [55]. Notable exceptions (i.e., background field methods that do not require masking) are recent deep learning approaches such as SHARQnet [169] and Total Field Inversion (which, however, requires a mask-like preconditioner) [170]. The performance of background field removal algorithms depends strongly on the mask and poor background field removal can negatively affect the quality of the reconstructed susceptibility maps [171,172]. Many dipole inversion algorithms use masks to exclude voxels with unreliable field values from the susceptibility computation or use masks for regularization [83,139,173]. Finally, susceptibility maps should be masked for display purposes to exclude streaks and spurious information outside the brain (see Section 7 on dipole inversion and Section 9 on presentation and publication).

Background field removal methods remove fields induced by all susceptibility sources outside the supplied brain mask. The accuracy of background field removal is lowest at the boundary of the ROI, such as the brain surface, and improves with increasing distance from the brain surface [55]. Although further erosion of the mask after BFR is not explicitly required, it may be employed in specific cases to eliminate residual artifacts at the boundary. Also, many background removal algorithms are not able to recover a reliable local field over the whole ROI mask [55], and further erosion is unavoidable. It should be noted that some background field removal algorithms (e.g., V-SHARP; see Section 6 below) will result in erosion of the brain mask and the resulting eroded mask should then be used in all subsequent operations and for display.

Holes inside the brain mask lead to the elimination of field contributions from the susceptibility sources within the holes during background field removal. Such holes can occur when thresholded (e.g., phase-based quality) maps are used to refine masking. For example, if a pathology (such as a hemorrhage or calcifications) creates unreliable phase data inside the brain, the affected region could be set to zero in the brain mask. Removing the field perturbations from susceptibility sources within the holes during the background elimination step will render these sources undetectable in the final susceptibility maps, which can be a significant problem in the clinical setting. Therefore, it is important that holes within the brain mask are filled before the mask is used for background field removal. Holes can be reintroduced in the mask used for dipole inversion as an effective way to prevent streaking artifacts from regions with unreliable phase values. This procedure has been included in some algorithms [139,174]. Since dipole inversion is a nonlocal operation, correct susceptibility values may be inferred inside relatively small holes excluded from the dipole inversion mask. To avoid streaking artifacts created by high contrast sources and pathologies, while preserving accurate susceptibility values inside the holes, some recent two-step approaches suggest performing reconstructions with and without holes and then merging both results [138,149,175,176]. This is useful to improve the accuracy of the reconstructions, and to characterize hemorrhages or calcifications.

Although some recent deep learning-based single-step QSM approaches have shown that explicit masking could be avoided [169,172,177–179], these methods require further study and validation to be considered for clinical applications.

## 5.2. Consensus Recommendations

The recommendations below are summarized in Figure 5, and differences between masks (Masks 1-4 in Fig. 5) are highlighted in Figure 6.

- 5.a. Create an initial brain mask (Mask 1) by applying a whole-brain segmentation tool (such as BET) to either the combined (sum of squares) or the first echo magnitude image. The goal of this initial mask is to remove air, skull and other tissues, while preserving cortical areas. Further refinement is performed in the following steps.
- 5.a. Create a mask of reliable phase values (Mask 2) by thresholding the phase quality map generated by the multi-echo combination method in Section 4. Multiply Mask 1 with Mask 2.
- 5.b. After multiplication, holes should be filled to obtain the mask to be used as an input to background field removal algorithms (Mask 3).
- 5.c. Holes from Mask 2 can be reintroduced to avoid streaking artifacts from unreliable phase data within the brain. For increased accuracy of susceptibility values inside pathological regions of low signal, e.g., hemorrhages and calcifications, mixing data from reconstructions with and without the holes can be performed.
- 5.d. The calculated susceptibility map should be multiplied by the mask used for background field removal (without holes; Mask 3) before display, reporting of susceptibility values or further analysis.

## 5.3. Additional Considerations

# 6. Background Field Removal

This section provides recommendations for the background field removal step.

## 6.1. Overview

In QSM, the background field is defined as the field generated by susceptibility sources outside a chosen ROI [55], in our case the brain mask (see previous section). In the brain, the background fields are generated by the tissue and air surrounding the brain. The susceptibility difference between brain tissue and air is approximately 9 ppm [76], which is almost two orders of magnitude larger than the naturally occurring susceptibility differences within the brain parenchyma. Therefore, background fields can be significantly larger than the tissue field in the brain, but not always are. Certain pathologies, such as hemorrhages, can create a tissue field that is similar in magnitude locally. The term *local field* is also often used in the literature for fields generated by tissue within the ROI, but since the field is a nonlocal property, we use the term *tissue field* here. Removal of the background field from the field map, $\Delta B_0$, allows focusing the inversion (see Section 7) on the spatial susceptibility variations located inside the ROI, which generate the so-called tissue field $\Delta B_t$ (Figure 7). When background fields are not completely removed from $\Delta B_0$, most dipole inversion methods will result in shadowing artifacts and/or experience a slow convergence rate.

Because of the spatial smoothness of the background field, spatial high-pass filtering has been a popular method to suppress background fields in the past. However, high-pass filtering also removes the low spatial frequency component, a major signal component, of the tissue field, which is not acceptable to QSM that requires quantitative accuracy of the corrected field maps. Newer methods that directly exploit the harmonic function property of background fields have replaced heuristic filtering methods [55]. From Maxwell's equations, it can be derived that the background field is a harmonic field, i.e., it satisfies the Laplace equation within the ROI. A harmonic field is completely determined when it is known on the region boundary. In other words, the solution of the Laplace equation in a region with a given boundary condition is unique [55,180].

The SHARP (Sophisticated Harmonic Artifact Reduction for Phase data) method [181] and variants thereof use the spherical mean value property of harmonic functions. This property implies that the average of a harmonic function over an arbitrary sphere centered at any location that fits within the region of interest is equal to the value of the harmonic function at that location. In practice, a radius is chosen that is somewhat large compared to the voxel size to overcome discretization effects. This means that the tissue field can only be computed for voxels that are at a distance equal to the chosen radius away from the boundary of the ROI. This limitation leads to an erosion of the region in which susceptibility can be computed (see also previous section). The most common variant of this method is V-SHARP [182], which involves multiple partial applications of SHARP with different radii to mitigate the practical implications of the erosion. V-SHARP yields background-corrected field values in the close vicinity of the ROI boundary but the values are not entirely accurate [55]. E-SHARP [183] and other variants of SHARP [184,185] overcome the remaining erosion of one voxel required for V-SHARP. Other variants like HARPERELLA [186] combine SHARP with phase unwrapping. In general, SHARP-based methods perform less well at the boundary of the region of interest [55]. In addition, SHARP-based methods include implicit low pass filtering due to the regularized deconvolution inherent in SHARP, which by itself removes slowly varying components [187].

The PDF (Projection onto Dipole Fields) method [188] finds an effective susceptibility distribution outside the region of interest that mimics the field inside that region. It uses the fact that the field generated by those outside sources are approximately orthogonal to those generated by local sources, allowing background field removal to be formulated as a noise weighted linear least squares problem. Because the orthogonality breaks down at the boundary of the region of interest, like SHARP, this method performs less well at the boundary [55,188].

The LBV (Laplacian Boundary Value) method [180] assumes that the field at the very boundary of the region of interest is entirely background field and determines a harmonic function that satisfies this boundary condition. An efficient algorithm has been introduced to solve this Laplacian boundary problem [180]. However, because LBV entirely relies on the field estimate at the mask boundary, it is sensitive to phase SNR at the boundary, rendering accurate masking particularly important. Often a small mask erosion is applied to remove low SNR voxels at the boundary. While LBV can perform better than PDF and SHARP in some situations[55], its performance has been observed to be highly dependent on the mask and on the quality of the field estimates at the mask boundary [189].

Residual background fields can be suppressed by combining methods, such as applying additional polynomial fitting or V-SHARP after LBV. V-SHARP and PDF typically do not require additional polynomial fitting. $B_1^+$ related contributions in the field map, $\Delta B_0$ (see Section 4 above) are not removed by background field removal methods, but these are avoided when using multi-echo data combined with field fitting as recommended above.

Susceptibility maps are dimensionless and are conventionally calculated and displayed in parts per million (ppm) (see Section 9.1.4). Assuming that the wrapped input phase was correctly scaled to (-$\pi$ to $\pi$) radians, the corresponding scaling can be done either before (on the tissue field map) or after (on the dipole inversion output) using the scale factor: $\Delta B_t \text{ (ppm)} = \frac{\Delta B_t \text{ (radians)} \cdot 10^6}{\gamma(\text{radians·T}^{-1}\text{·s}^{-1}) \cdot B_0(\text{T}) \cdot \Delta TE(\text{s})}$. When scaling is performed after, care has to be taken to adjust the default regularization parameter when using a total variation based dipole inversion method, as the regularization term scales linearly.

## 6.2. Consensus Recommendations

    6.a.    Use V-SHARP to achieve good results in many situations, as it is less sensitive to imperfections in brain masking. This comes at a cost of a one-voxel erosion of the brain mask used for dipole inversion (Mask 4 in Fig. 5) at the brain surface and reduced accuracy at the edge of the brain.

    6.b.    When whole brain mapping (including the cortex and superficial veins) is desired, use PDF. This method will be slightly more accurate throughout the brain. PDF requires a good brain mask.

    6.c.    Depending on the application, tissue field quality, i.e., the phase SNR especially near the boundary, must be balanced against mask erosion.

## 6.3. Additional Considerations

Single step [112,170,190–192] or total field inversion methods fit the susceptibility directly to the field map, $\Delta B_0$, or even wrapped phase images. These are currently popular for applications of QSM outside the brain but are still under ongoing development to ensure robustness. Residual background field has been tackled for dipole inversion using weak harmonics modeling [136]. Finally, deep learning [52] has found application in background field removal as well and is the subject of ongoing development.

# 7. Dipole Inversion

This section provides recommendations for the field-to-susceptibility inversion step (Fig. 8), which derives from the tissue field map, $\Delta B_t$ (with background fields removed; see previous section), a map that is tissue magnetic susceptibility $\chi(r)$ (up to a reference value [193,194], see Section 8).

## 7.1 Overview

While susceptibility, $\chi(\boldsymbol{r})$, is a local tissue property, the field is a summation of weighted contributions from the distribution of magnetic susceptibility in all space. Mathematically, this summation can be described as a convolution ($*$) of the susceptibility with the unit dipole kernel $d(\boldsymbol{r}) = \frac{1}{4\pi}\frac{3cos^2\theta - 1}{r^3}$ [77–79]:

$$\Delta B(\boldsymbol{r}) = d(\boldsymbol{r}) * \chi(\boldsymbol{r}). \qquad [1]$$

Convolution corresponds to multiplication in the spatial frequency domain, which facilitates its fast calculation and is used in most QSM implementations [77–79] to accelerate computations. The inversion step performs a deconvolution using the dipole kernel *d(r),* which reveals the local tissue susceptibility within the region of interest, $\chi_t(\boldsymbol{r})$, from the background-corrected tissue field, $\Delta B_t(\boldsymbol{r})$: $\Delta B_t(\boldsymbol{r}) *^{-1} d(\boldsymbol{r}) = \chi_t(\boldsymbol{r})$. However, the dipole kernel value is zero at and very small near the cone surface of the magic angle ($54.7^0$) relative to the direction of the main magnetic field, making this deconvolution a poorly conditioned inverse problem [75,84,195–197]. The measured tissue field, $\Delta B_t$, contains deviations from perfect dipole patterns, particularly in regions with small magnitude signal due to lack of water protons (field detectors) or rapid signal decay (largely caused by inhomogeneous fields). While the regions with most extreme deviations are usually eliminated through masking (see Section 5 above), remaining dipole deviations in the estimated field can cause deconvolution errors in the calculated susceptibility map reconstruction, manifesting as streaking and shadowing artifacts [53,198]. Additional information about the unknown susceptibility map, $\chi(\boldsymbol{r})$, can be incorporated into the solution through regularization to suppress streaking and shadowing artifacts in the solution [25,59,83,84,112,141,173,182,199,200]. An optimization approach for incorporating this additional information can be formulated according to Bayesian inference, which is the following minimization problem when approximating the noise in the field as Gaussian:

$$\chi(\boldsymbol{r}) = \mathrm{argmin}_{\chi(\boldsymbol{r})} \|w(\Delta B_t - d * \chi)\|_2^2 + \lambda R(\chi). \qquad [2]$$

Here the first term is the data fidelity term with spatially varying noise weighting $w$ and the second term, $R(\chi)$, is the regularization term with $\lambda$ as regularization strength [83,141]. The minimization problem is iteratively solved with the number of iterations determined by the desired convergence level. The optimal regularization strength ($\lambda$) depends on anatomy, susceptibility contrast, and SNR, and should be optimized to balance artifact suppression and image sharpness in each imaging protocol and application by varying $\lambda$ using, e.g., the L-curve method [201,202].

Various regularization strategies have been developed for the inverse problem in QSM [25,53,60,83,84,136,141,173,182,192,197,199–201,203–207]. Available QSM software packages include FAst Nonlinear Susceptibility Inversion (FANSI) [173] (https://gitlab.com/cmilovic/FANSI-toolbox), Morphology Enabled Dipole Inversion (MEDI) [201,208] (http://pre.weill.cornell.edu/mri/pages/qsm.html), and STI Suite [209] (https://people.eecs.berkeley.edu/~chunlei.liu/software.html).

Total variation regularization has performed favorably in the two QSM reconstruction challenges [60,63]. Both FANSI and MEDI provide specific implementations of sparsity regularization with openly accessible source code. Specific implementation examples including zero-referencing to the cerebrospinal fluid (CSF) may be included as an extra regularization that provides the CSF-uniformity verification on the QSM output with an additional benefit of further reducing streaking and shadowing artifacts, [193,210] but are associated with other limitation as discussed in the next section.

The simple sparse regularizer using L1 norm of the gradient (i.e., the total variation, TV) is a standard approach for brain QSM, as exemplified in MEDI, one of the most popular algorithms. The performance of the TV approach for brain QSM in terms of accuracy and robustness was well established in the 2019 QSM Reconstruction Challenge, particularly in the presence of strong susceptibility sources, such as hemorrhages or calcifications, using nonlinear forward signal modeling [60,63]. Recent developments in deep learning based QSM reconstruction represent an exciting avenue for improving QSM performance [52,179,211,212]. However, while there are instances where these methods yield better results than classical methods, the performance of these methods did not reach those of the best classical methods in the QSM challenges potentially due to generalization issues from limited training data [60,63].

Some algorithms do not incorporate spatial constraints for suppressing streaking and shadowing artifacts but explicitly modify the dipole kernel instead [53,198], for example, the thresholded k-space division [192,197,213]. Implicit regularizations based on the number of iterations may work, but these methods have limited denoising capabilities and may be less robust than the sparsity regularization optimization approach [59,205,214].

## 7.2. Consensus Recommendation

7.a. Use an optimization approach for dipole inversion with a sparsity type regularization that is commonly used in compressed sensing [53]. Specific sparsity types include L1-norm, total variation, and generalized total variation, which likely provide similar outcomes. Future algorithm developments and evaluations are needed to provide a more specific consensus on the sparsity type.

7.b. Use the default sparsity type, regularization strength and number of iterations in a QSM software, such as the processing pipelines recommended here (Supplementary Materials II), including FANSI, STI Suite, and MEDI, where these default parameters have been optimized for common brain protocols. If the acquisition protocol recommended here (Supplementary Materials I, Section S1.5) is substantially altered, researchers should perform an L-curve optimization or other method on at least one typical case with the specific study protocol to finetune the regularization strength and iteration number and then fix these parameters for the same protocol.

## 7.3. Additional Considerations

There may be streaking artifacts coming from strong susceptibility sources near borders and within the brain interior region. Major causes include the breakdown of the Gaussian noise assumption and other errors in the determined field [139]. These artifacts may be suppressed using methods such as masking out or reducing the weight of less trustworthy voxels in the optimization [139]. The border streaking can be removed by improving the brain mask [149,215,216]. The interior streaking can be reduced using techniques to improve convergence such as preconditioning [170,217], and using in-painting techniques to compensate for field errors such as MERIT [139] and $L_1$ data fidelity [174].

There may be shadowing artifacts coming from residual background fields. This shadowing can be reduced by improving background field removal such as harmonic incompatibility removal [136,218] and by suppressing slowly varying spatial frequency components through regularization [219] or preconditioning [53,170].

# 8. Analysis of Susceptibility Maps

This section provides recommendations for quality control and referencing of susceptibility maps, the quantification of susceptibility values, and the visualization of brain structures on susceptibility maps and derived contrasts in the context of clinical research performing group studies. The physical background for the consensus recommendations is briefly summarized. Possible tools to facilitate susceptibility quantification of brain structures and lesions as well as for group analyses are provided in Supplementary Materials I, Section S1.6. Figure 9 summarizes the recommendations of this section.

## 8.1. Overview

### 8.1.1. Quality control

Only a few tools for fully automatic QSM calculation and evaluation directly from scanner DICOM data exist to date that perform all steps outlined in Sections 2 to 8 [149,220]. Some of these tools may not be suitable for all possible QSM applications due to assumptions on patient cohorts of the implemented mask generation algorithms (see Section 5 above) or due to the need to adjust reconstruction parameters depending on the data (see Section 7 on dipole inversion). Most QSM applications still require multiple processing steps, which can result in error amplification/propagation or inconsistencies between steps, rendering QSM workflows prone to i) reconstruction artifacts (a list of common reconstruction artifacts is provided in Table 1) and ii) calculation errors of region-specific susceptibility values. Particularly, the use of one or several masks to exclude unreliable phase data and for background field removal during QSM calculation can result in missing areas in computed susceptibility maps especially close to air-tissue interfaces. When voxels in those regions are not properly excluded in subsequent analyses of the susceptibility maps (e.g., in ROI-based analyses), regional mean values may be biased by these erroneously included zero-valued voxels in the susceptibility mean value calculations. The issue can be resolved by incorporating the eroded background-correction mask (Mask 4 without holes in Fig. 5) in the ROI masks. Deviation from the radiological orientation (right-left flip) in the final susceptibility maps (see also Section 3.1) can be a potential issue arising from the combination of different toolboxes when using other tools than those recommended here or as a result of erroneous use. These flips can have detrimental consequences in the clinical setting but can be revealed relatively easily comparing brain features between QSM and the original GRE magnitude images on the scanner console, especially when the subject's head was tilted to the right or left (about the H-F, A-P or both of these axes) for test purposes (see Section 3.1), a step that should always be done if custom pipelines are used.

### 8.1.2. Referencing and choice of reference region

QSM can only assess relative susceptibility differences between tissues as phase data reflect field distortions due to these underlying spatial susceptibility differences. Susceptibility values are therefore given up to a reference [56,194]. To obtain susceptibility values that are comparable between repeated measurements, subjects, and scanners, consistent referencing of susceptibility maps is required. In QSM, internal reference regions are used. External reference regions are not generally used because it is not currently possible to measure phase differences between disconnected spatial regions separated by noise and perform consistent background field removal for both the brain tissue and the external reference region. The ideal choice of a reference region for brain QSM is still under debate [194]. Different regions used in the literature come with certain advantages and disadvantages and will lead to different susceptibility values in the resulting susceptibility map. For example, assuming an ROI's average susceptibility value is 0.010 ppm when computed from a susceptibility map referenced to the whole brain (with assumed mean susceptibility of whole brain -0.001 ppm), this ROI susceptibility value will be 0.008 ppm when computed from a susceptibility map referenced to the CSF (assumed mean susceptibility of CSF 0.001 ppm).

In the case of widespread pathology such as in multiple sclerosis or Alzheimer's disease, there might not be an ideal choice of reference region. Larger reference regions are generally advantageous over small-sized regions, which are more affected by potential local lesions, reconstruction inhomogeneities and other artifacts (less averaging), which are then propagated to all other regions in the map by the referencing process. This issue reduces statistical power and therefore 3D segmentation of reference regions is advisable to include a greater number of voxels. Consequently, whole-brain referencing is considered stable (largest possible mask) and reproducible (whole-brain mask readily available in all reconstruction pipelines).

The dependence of white matter apparent susceptibility on the fiber orientation with respect to the main magnetic field due to the geometry and complex microstructure of white matter fiber bundles [102,104,146,221,222] (see also Section 9.1.6) can be a source of additional variability when using a reference that includes white matter regions. Another challenge of referencing is that pathology or effects of age alter white and gray matter integrity, specifically myelination and brain iron levels, especially in deep gray nuclei [223–228]. In the case of widespread pathology when no ideal reference region exists, two reference regions could be used to evaluate if the choice of reference region affects the study results. If, for example, whole brain and CSF were used as reference leading to the same significant differences between patients and controls, the results can be assumes with greater confidence to originate from the presence of pathology, instead of being an artifact from susceptibility referencing [229].

For local pathology, the use of contralateral or surrounding tissues as reference is an effective strategy to avoid introducing artificial susceptibility differences due to using a reference region affected by pathology.

Table 2 lists advantages and disadvantages of common reference regions. More details on referencing can be found in dedicated literature [193,194]. We recommend referencing with regions that are commonly used in the literature. In addition, we recommend that studies report the mean and standard deviation of the susceptibility (after referencing) in other regions that are or have been widely used for referencing along with their hypothesis-driven regions of interest. This approach will promote reproducible research as it facilitates the comparison of susceptibility values between studies and enables post hoc re-referencing for meta-analyses. While no normative susceptibility values exist, literature values [230] can serve as precedence reference when comparable subjects are studied, e.g. healthy controls of similar age.

## 8.1.3. Effect of segmentation on susceptibility quantification (iron, white matter changes, lesions, vessels, oxygenation)

An accurate segmentation of ROIs is essential to uncover subtle changes in regional susceptibility values that might indicate pathology, or to establish normative values. While manual segmentation of regions by multiple expert readers is the gold standard for quantification of regional susceptibility values, this strategy is very time-consuming and therefore not feasible in larger studies. Many available automated neuroimaging segmentation tools are optimized for use with $T_1$-weighted images or require $T_1$-weighted input data [231]. However, when using these methods for the analysis of susceptibility maps, the segmentation and registration accuracy in many structures of interest (e.g., basal ganglia) can depend on $T_1$ contrast [232], which is also affected by tissue iron [233,234], and the generally low visibility of some deep gray matter regions on $T_1$-weighted images [235] (depending on sequence parameters). Consequently, ROI-based methods that rely solely on $T_1$-weighted contrast may be biased and suffer from inaccuracies. Previously, it has been shown that the use of a QSM or hybrid QSM-$T_1$-weighted contrasts for template generation improves atlas and voxel-based analyses [30,31]. Therefore, using multi-contrast segmentation can be considered the best approach to avoid template bias [236]. A list of recommended tools can be found in the Supplementary Materials I, Section S1.6. Partial volume effects might strongly affect susceptibility quantification both for voxel-based and ROI-based analyses, especially for small structures with relatively high susceptibility values such as veins [237]. This could be corrected for by eroding of ROIs [101], only using high susceptibility voxels (in case of positive susceptibility) [238] or using a partial volume map for correction [239].

## 8.2. Consensus Recommendations

8.a. When ROIs are affected by artifacts, exclude data by automated detection of outliers or outlier regions, use of image quality measures or visual inspection.

8.b. Ensure that analysis methods do not include voxels of the susceptibility map with unreliable values, e.g., that lie outside of the eroded background field removal mask (see Section 5 above; Mask 4 without holes in Fig. 5).

8.c. Always reference susceptibility maps to an internal reference region before performing further analyses.

8.d. When choosing a reference region, consider the study design, influence of pathology, how pathology could bias the study findings and discuss accordingly. For widespread pathology, cross-checking results using two different reference regions (e.g., whole brain and CSF) can be considered safe to exclude bias.

8.e. Segment reference regions in 3D.

8.f. Always include commonly used reference regions in your analysis and report mean and standard deviation in these regions along with those in other ROIs.

8.g. Consider incorporating QSM contrast in ROI segmentation or ensure that $T_1$w-based methods are accurate.

## 8.3. Additional Considerations

# 9. Presentation and Publication

## 9.1. Overview

The purpose of the recommendations in this section is to facilitate the interpretation and replicability of future findings with QSM, future meta-analyses, and the comparison among studies. To this end, the general recommendation is to report as much information as possible regarding:

1. Data acquisition (hardware and scan parameters);
2. Reconstruction pipeline and analysis procedure; and
3. Results.

Depending on the study and on the journal in which the study will be published, the degree of information detail that can be reported may vary. The members of the QSM Consensus Organization Committee asked themselves, through a multiple-choice grid form, whether each information entity relevant for QSM should be reported *always* (*a*) or only *depending* (*d*) on the study and on the journal, or if it is *unnecessary* (*u*) to report the entity. Each item in the poll was assigned a score $S = (A + 0.5D)/(A + D + U)$, where *A*, *D* and *U* are the number of *a*, *d* and *u* responses collected for that item, respectively. The reporting of specific items was considered essential if there was unanimous consensus in reporting them among the authors ($S = 1$). Items that were not considered essential were assigned a "traffic light ranking" (green for $0.75 \leq S \leq 1$, orange for $0.5 \leq S \leq 0.75$, and red for $S \leq 0.5$). Standardized tables are provided to facilitate the reporting of a broad set of items. For the purpose of availability, unless there is limited space for particular journals, we recommend that items with S>0.5 be reported as other investigators may need this information. It is considered essential to report these items if they vary within the same study (e.g., if different scanners or different software releases are used within the same study). Potential limitations and confounds should always be discussed. The last part of this section reviews some important aspects that should always be considered when interpreting and presenting QSM findings in scientific papers.

### 9.1.1. Acquisition hardware

Ideally, the acquisition hardware is described in one sentence reporting the scanner field strength, model, vendor, software release version, and type of coil used (including the number of channels). Table 3 provides an overview of consensus recommendations pertaining to acquisition hardware.

### 9.1.2. Acquisition sequence type and parameters

The QSM Consensus Organization Committee considered it as essential to indicate the acquisition sequence type (e.g. GRE, as recommended in this paper, or EPI; specify if the sequence is 3D or 2D) and several acquisition parameters including number of echoes, TEs, TR, flip angle, bandwidth, resolution and scan duration. Table 4 provides an overview of consensus recommendations pertaining to acquisition sequence type and parameters.

### 9.1.3. Reconstruction pipeline and analysis

It is considered essential to describe the toolbox and reconstruction pipeline and list the algorithms used. The numerical values of parameters used should be listed, even if they were the default parameters. Table 5 provides an overview of consensus recommendations pertaining to the reconstruction and analysis pipelines.

### 9.1.4. Displaying figures

When displaying quantitative susceptibility maps, do not use rainbow, jet, or similar types of non-linear colormaps, which introduce the erroneous perception of artificial edges in some parts of the range, hide existing edges in other parts of the range, and lack intuitive perceptual ordering [240–242]. In the absence of a motivation for doing otherwise in particular studies, the use of a linear, perceptually uniform colormap should be preferred; the use of a linear gray-scale map enables consistency with the vast majority of the published literature. This applies also to phase data, to enable a clear representation of phase wraps and/or possible errors such as open-ended fringe lines. Contrast windowing should be adjusted to avoid saturation of relevant brain areas (i.e., completely black or white appearance). A typical window adapted to healthy brain QSM is [-0.2, +0.2] ppm. The windowing should always be reported, by either using an intensity bar or writing the information in the figure caption. Susceptibility maps should be displayed through the eroded mask used for background field removal (see Section 5), to avoid representation of artifactual data outside the brain.

### 9.1.5. Sample paragraphs

Representative paragraphs describing data acquisition, processing and QSM calculation in a scientific paper are provided in the following. The description refers to the images shown in Figure 8.

*"Data were acquired on a 3 T scanner (Prisma Fit, Siemens Healthcare, Erlangen, Germany; VE11B) using the built-in whole-body RF transmit coil and a 64-channel receive-only head/neck coil. The acquisition sequence was a 3D GRE multi-echo with pure axial orientation with the following scanning parameters: TR = 33 ms, 5 monopolar echoes acquired at $TE_1 : \Delta TE : TE_5 =$ 5.25 : 5.83 : 28.57 ms, flow compensation for the first echo in the readout (AP) and 'slice' encoding (HF) direction, FA = 15°, pixel bandwidth = 220 Hz, elliptical k-space shutter, covering a field of view (FOV) of 256(AP)×176(LR)×144(HF) $mm^3$ with matrix size=256×176×144, resulting in isotropic voxels of size 1$mm^3$, with GRAPPA acceleration factor=2 in the phase encoding (LR) direction. Scan duration was 6 minutes 34 seconds. The full QSM reconstruction was performed in Matlab R2021a (MathWorks, Natick, MA, USA) using the SEPIA toolbox [73] (v1.2.2.4) for integration of the various processing steps described hereafter. ROMEO total field calculation [134] (v3.5.6) was used for echo phase combination. Brain masking was obtained with FSL BET [158] on the first-echo magnitude image, using default settings. V-SHARP [182] with spherical mean value filtering sizes from 12 mm to 1 mm was used for background field removal. Quantitative susceptibility maps were obtained using FANSI [173] (v3) with gradient L1 penalty of 0.0005 and gradient consistency weight of 0.05. Susceptibility values are expressed in parts per million (ppm) and have been referenced to the average susceptibility in the brain mask (imposed to zero by the adopted processing pipeline)."*

A complete description of the QSM calculation pipeline can be found in the Supplementary Materials II.

### 9.1.6. Interpretation of results

Potential limitations and confounds related to QSM should always be taken into account.

A potential confound that can affect the extraction of quantitative susceptibility values from MRI phase/frequency data arises from the fact that, even for a uniform voxel-averaged susceptibility distribution, the apparent field measured in a voxel depends on the subvoxel distribution and visibility of water protons (the sensors of the MRI signal) around susceptibility perturbers such as iron and myelin. This can lead to a phase shift resulting from the biased sampling of the fields generated by perturbers when sensor and perturber distributions spatially correlate and such correlation is anisotropic [243,244]. An example is water in and around myelinated fibers, whose anisotropic distribution leads to a fiber orientation-dependent shift in apparent frequency which can exceed 10 ppb [104,144,146,221,243,244]. In addition, substantial $T_1$, MT, or $T_2^*$ weighting may differentially affect water visibility in the various water compartments and render apparent frequency shifts dependent (in a non-linear manner) on TR or TE [142,143,145]. Especially above 3T, QSM values within and around fibers that run perpendicular to $B_0$ should be interpreted with caution. For example, pathological changes in myelin structure but not myelin content in such fiber bundles may lead to QSM changes without actual changes in tissue susceptibility.

Another point of caution with interpretation of QSM are inaccuracies near the edge of the regions selected for the analysis (see Section 5). A notable example are areas near the surface of the brain, where phase data is unreliable (due to, e.g., the prevalence of paramagnetic blood in pial veins), or unavailable (due to the lack of signal in skull), or the tissue phase was partially removed in the background field removal step. Because of this, QSM values in some of cortical grey matter may be incorrect or have reduced spatial contrast and resolution.

Lastly, it should be realized that, when strong regularizations or prior information are used in QSM dipole inversion, potential smoothing and spatial resolution loss may occur or new features may be added [201,245,246] (see Section 7.1). Some anatomical detail, visible in phase or magnitude GRE data, may therefore be lost in the QSM.

## 9.2. Consensus Recommendations

9.a. Always report at least the essential information regarding the acquisition hardware (Table 3), acquisition sequence type and parameters (Table 4), reconstruction pipeline and analysis (Table 5).

9.b. Representative susceptibility maps and the underlying background-field corrected phase images should be shown in all articles.

To facilitate documenting the reconstruction pipeline, we encourage software developers to enable printing out the values of all relevant parameters in Tables 4 and 5 (including default parameters) and provide suggested descriptions of their toolboxes, which users can re-utilize in their publications.

## 9.3. Additional Considerations

# 10. Summary and Conclusion

This consensus paper has been developed by the QSM Consensus Paper Committee with consideration of suggestions from the whole QSM research community (see Acknowledgements and Supplementary Materials I, Section S1.3). The paper provides recommendations for all steps essential in setting up a successful QSM study in a clinical research setting. The recommendations, intended for a robust but not necessarily state-of-the-art QSM, are based on the current understanding (as of 2023) and should be updated as the QSM field progresses.

In summary, we recommend that data be acquired using a monopolar 3D multi-echo GRE sequence, that phase images be saved and exported in DICOM format and unwrapped using a quantitative approach. Echoes should be combined before background removal, and a brain mask created using a brain extraction tool with the incorporation of phase-quality based masking. Background fields within the brain mask should be removed using a SHARP-based or PDF technique and the optimization approach to dipole inversion should be employed with a sparsity type regularization. Susceptibility values should be measured relative to a specified reference, including the common reference region of whole brain as a region of interest in the analysis, and QSM results should be reported with – as a minimum – the acquisition and processing specifications listed in the final section.

The recommended steps for data acquisition, data preparation and post processing are intended to provide a uniform robust reference starting point for a brain-focused QSM study performed with a clinical scanner. Specialty applications such as the depiction of small structures might require spatial resolutions higher than recommended [247]. In this regard, limitations and further considerations are included in each section, but thorough testing of the processing pipeline is recommended before starting a large patient study.

We hope that the recommendations here will enable many medical research centers to perform comparable QSM studies on scanners from different vendors, and that the standardized acquisition protocols and the processing pipeline provided along with this article will facilitate these studies (see Supplementary Materials I, Section S1.5. and Supplementary Materials II). As more clinical QSM studies are performed, analyzed, and presented in scientific publications, and current and future technical innovations become mature, these QSM recommendations will need to be updated.

# Acknowledgements

We thank the following individuals for their significant contribution to the development of the initial consensus statements during the study group review phase: Steffen Bollmann (The University of Queensland), Marta Lancione (IRCCS Stella Maris Foundation), Jakob Meineke (Philips Research Hamburg), Xi Peng, Ludovic de Rochefort (Aix-Marseille University), Mathieu Santin (L'Institut du Cerveau et de la Moelle Épinière), Salil Soman (Harvard Medical School).

We thank the following individuals for their significant contribution to the improvement of the manuscript at and following the presentation at the 2022 Joint Workshop on MR phase, magnetic susceptibility and electrical properties mapping held in Lucca, Italy on October 16-19, 2022. Researchers providing general advice on consensus recommendations: Xavier Golay (University College London); MRI physicists supporting clinicians: Wiebke Nordhøy (Oslo University Hospital); MRI physicists working at manufacturers: Brian Burns (GE Healthcare), Kim van de Ven (Philips); Neurologists: Nicolás Crossley K. (The Pontifical Catholic University of Chile).

In addition, we thank the following ISMRM EMPT study group members, industry representatives, and clinicians who provided substantial feedback during the November/December 2022 manuscript review phase: Antje Bishof (University Hospital Münster), Yoshitaka Bito (Industry; Fujifilm), Nicolas Crossley (Pontificia Universidad Catolica de Chile), Andreas Deistung (University Hospital Halle), Cristina Granziera (University of Basel), Mark E. Haacke (Wayne State University), Marta Lancione (IRCCS Stella Maris Foundation), Emelie Lind (Lund University); Anna Lundberg (Lund University); Jie Luo (Shanghai Jiao Tong University), Jakob Meineke (Philips Research Hamburg); Kieran O'Brian (Industry; Siemens Healthineers), Alexander Rauscher (University of British Columbia), Daniel Sodickson (NYU Langone Health), Hongfu Sun (University of Queensland), Anil Man Tuladhar (Radboud University Medical Center), Anja van der Kolk (Donders Institute), Kim van de Ven (Industry; Philips); Alan Wilman (University of Alberta), Ronnie Wirestam (Lund University); Yongquan Ye (Industry; United Imaging); Xiangzhi Zhou (Mayo Clinic).

This publication was supported by the National Institute of Neurological Disorders and Stroke of the National Institutes of Health under Award Number R01 NS114227 (F.S.), R01 NS105144, and R01 NS095562 (Y.W.), the National Center for Advancing Translational Sciences of the National Institutes of Health under Award Number UL1 TR001412 (F.S.), the National Institute of Biomedical Imaging and Bioengineering of the National Institutes of Health under Award

Numbers R01 EB032378, R01 EB028797, R03 EB031175, P41 EB030006, U01 EB026996 (B.B.), U01 EB025162 (B.B., C.L.), and P41 EB031771 (P.V.Z, X.L.), the National Institute on Aging of the National Institutes of Health under Award Numbers R01 AG063842 (X.L.) and R01 AG070826 (C.L.), and the National Institute of Mental Health of the National Institutes of Health under Award Number R01 MH127104 (C.L.). The content is solely the responsibility of the authors and does not necessarily represent the official views of the National Institutes of Health. S.R. was supported by the Austrian Science Fund (FWF): 31452 and the European Union's Horizon 2020 research and innovation programme under the Marie Skłodowska-Curie grant agreement No. 794298. K.S. is supported by European Research Council consolidator grant DiSCo MRI SFN 770939. M.C. is supported by the Italian Ministry of Health (grant RC and 5×1000 voluntary contributions)

## CRediT authorship contribution statement

**Berkin Bilgic**: Conceptualization, Methodology, Writing – Original Draft Preparation, Writing – Review & Editing. **Mauro Costagli**: Conceptualization, Methodology, Visualization, Writing – Original Draft Preparation (lead), Writing – Review & Editing. **Jeff Duyn:** Writing – Review & Editing. **Christian Langkammer**: Conceptualization, Methodology, Writing – Original Draft Preparation, Writing – Review & Editing. **Jongho Lee**: Conceptualization, Methodology, Visualization, Writing – Original Draft Preparation (lead), Writing – Review & Editing. **Xu Li**: Conceptualization, Methodology, Visualization, Writing – Original Draft Preparation (lead), Writing – Review & Editing. **Chunlei Liu:** Conceptualization, Methodology, Writing – Original Draft Preparation, Writing – Review & Editing. **José Marques:** Conceptualization, Methodology, Resources, Software, Visualization, Writing – Original Draft Preparation (lead), Writing – Review & Editing. **Carlos Milovic:** Conceptualization, Methodology, Visualization, Writing – Original Draft Preparation (lead), Writing – Review & Editing. **Simon Robinson:** Conceptualization, Methodology, Visualization, Writing – Original Draft Preparation (lead), Writing – Review & Editing. **Ferdinand Schweser:** Conceptualization, Methodology, Project Administration (lead), Supervision (lead), Writing – Original Draft Preparation, Writing – Review & Editing (lead). **Kwok-Shing Chan:** Conceptualization, Data Curation, Formal Analysis, Methodology, Software (lead). **Karin Shmueli:** Conceptualization, Methodology, Writing – Original Draft Preparation, Writing – Review & Editing. **Pascal Spincemaille:** Conceptualization, Methodology, Visualization, Writing – Original Draft Preparation (lead), Writing – Review & Editing. **Sina Straub:** Conceptualization, Methodology, Visualization, Writing – Original Draft Preparation (lead), Writing – Review & Editing. **Peter van Zijl:** Conceptualization, Writing – Review & Editing. **Yi Wang:** Conceptualization, Methodology, Resources, Supervision, Visualization, Writing – Original Draft Preparation (lead), Writing – Review & Editing

257. Zivadinov R, Tavazzi E, Bergsland N, et al. Brain Iron at Quantitative MRI Is Associated with Disability in Multiple Sclerosis. *Radiology*. 2018;289(2):487-496. doi:10.1148/radiol.2018180136

# Figures

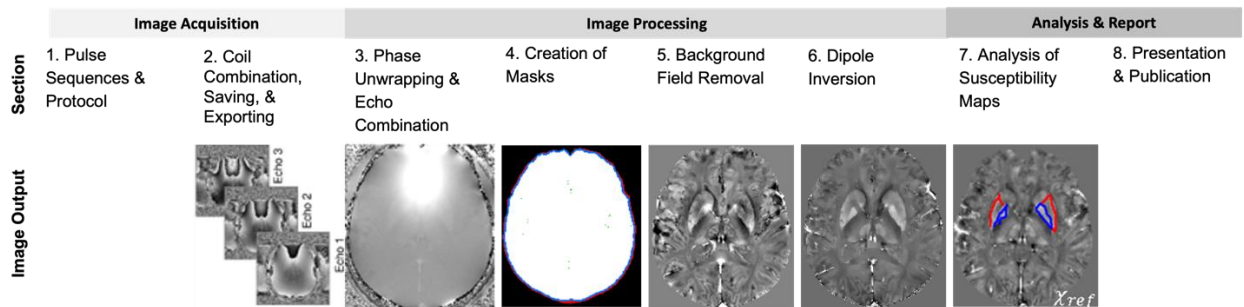

**Figure 1**: This consensus paper comprises eight sections. The first two sections cover image acquisition: 1) pulse sequences and protocol and 2) coil combination, saving, and exporting. The next four sections cover image processing: 3) phase unwrapping & echo combination, 4) creation of masks, 5) background field removal, and 6) dipole inversion. The last two sections cover analysis and presentation in scientific publications: 7) analysis of susceptibility maps, and 8) presentation and publication. The image output from each section is further detailed in the corresponding section.

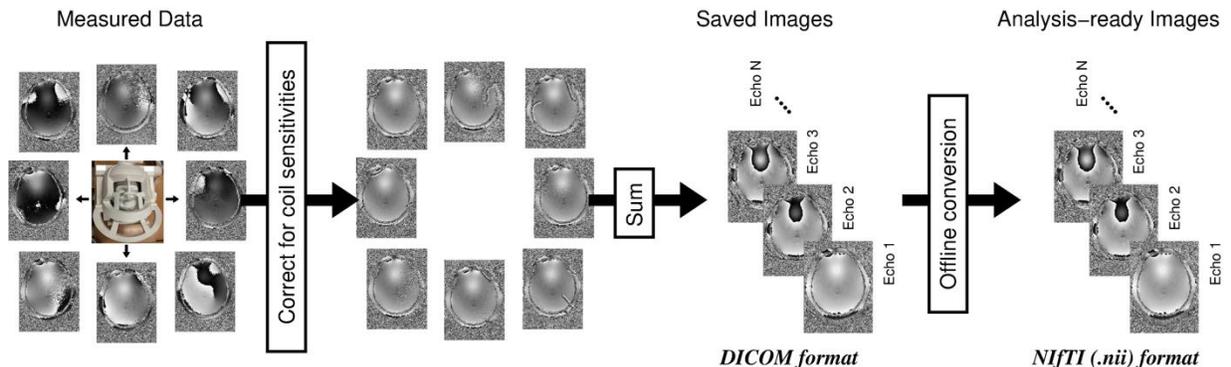

**Figure 2:** Steps in coil combination, saving, and exporting (illustrated for eight example coils from the 64-channel array). Each of the coils generates a phase image (left), which is modified by the coil sensitivity and other terms which make up the initial phase. The initial phase is removed using methods detailed in the text and referenced publications (center left) and phase images are combined in the manufacturer's reconstruction and saved for export in DICOM format (center right). QSM analysis software may require the DICOM data to be converted, offline, to NIfTI format (right). Images shown were acquired at 3 T (Siemens Healthineers, Erlangen, Germany; Prisma Fit, VE11C) with a head/neck 64 channel coil and the recommended multi-echo GRE sequence ($TE_1$=5.25ms; echo spacing=5.83ms; 5 echoes) using monopolar readout. Additional details are reported in Section 9.1.5. The imaging data and a description of the acquisition protocol may be found in Supplementary Materials II.

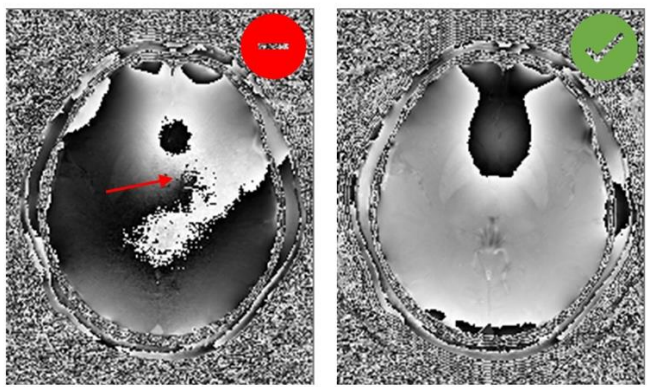

**Figure 3**: Some scanner manufacturers' options for processing and saving phase images (like "Sum of Squares") do not remove coil sensitivities. This may be apparent in the combined phase images having open-ended fringe lines (left). Wraps in phase images combined with the recommended methods are quite symmetric across the brain mid-line (right), and (like contours on a topographic map) either begin and end at the edge of the brain tissues, or form closed loops within the brain.

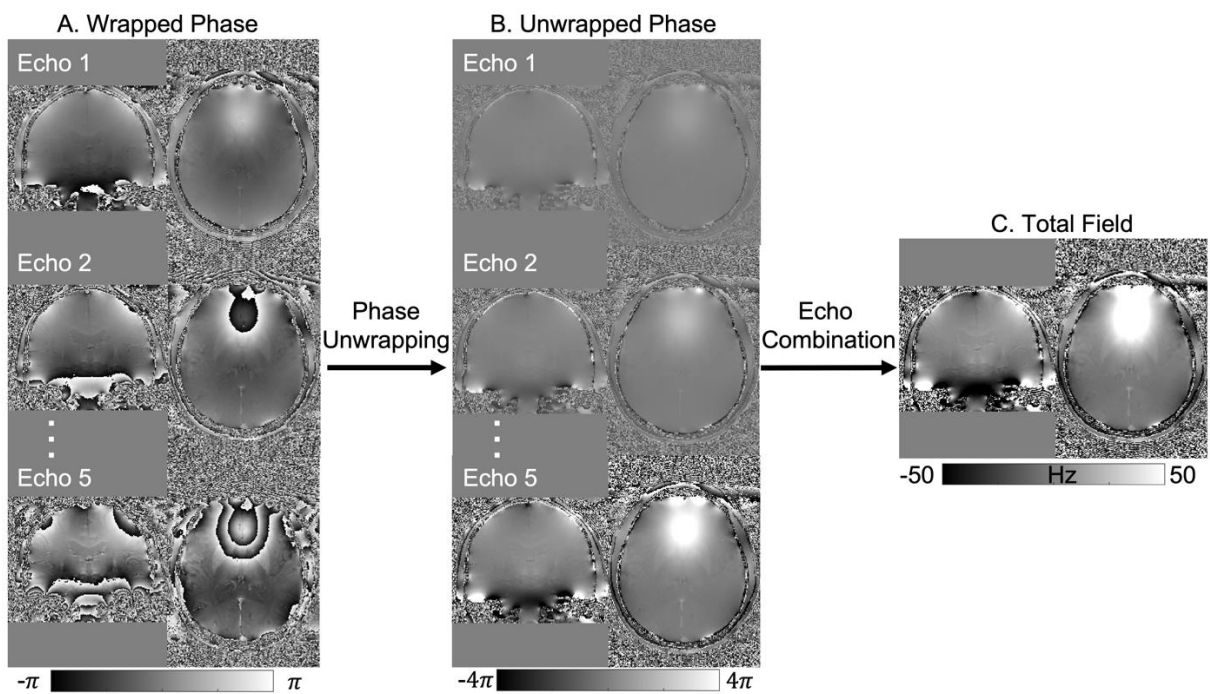

**Figure 4.** A) Example wrapped phase images at different echo times after proper coil combination (same images as shown on the right-hand side of Figure 2). More phase wraps can be observed at a later echo (bottom). B) Example unwrapped phase images (using ROMEO template phase unwrapping with MCPC3D-S phase offset correction). C) Total field estimation after echo combination using weighted echo averaging.

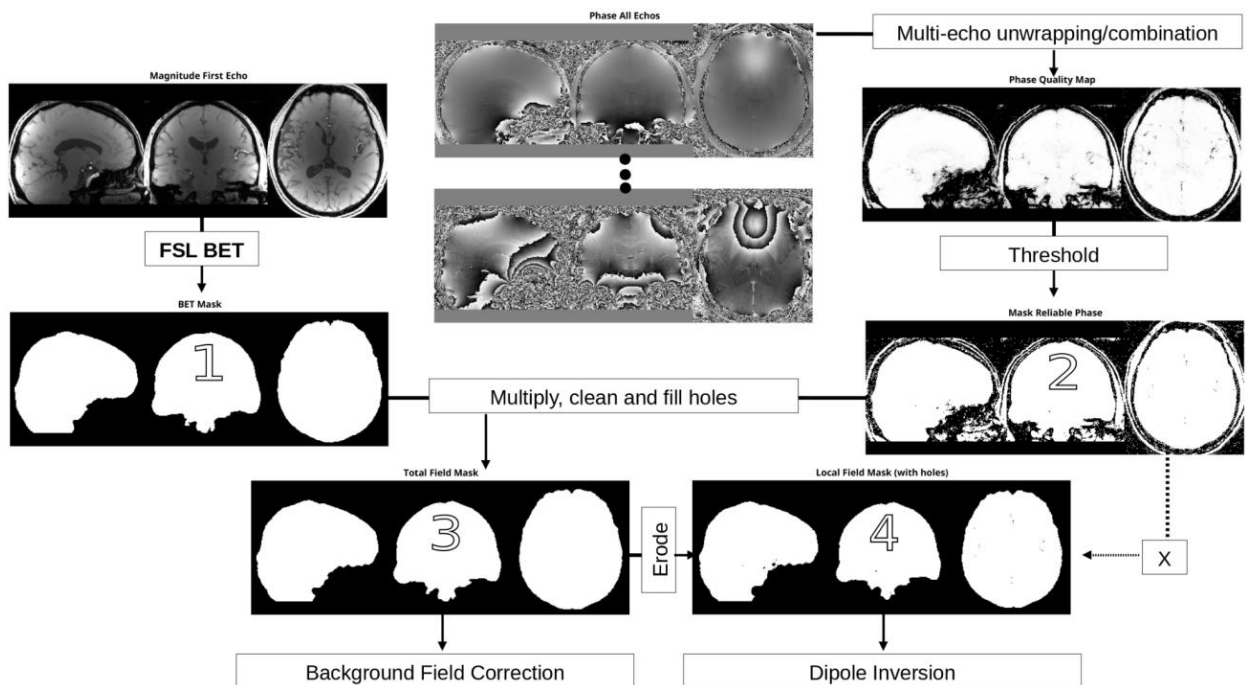

**Figure 5:** Block diagram of the masking stages. 1) Create an initial mask using FSL BET (Mask 1). 2) Threshold a phase-based quality map to create a reliable phase mask of reliable phase values (Mask 2). 3) Multiply Mask 1 with Mask 2 and fill holes for background field removal. 4) Erode by one or two voxels according to the output of the background field removal algorithm (and optionally reintroduce holes) for dipole inversion. Use Mask 4 without holes filled in for display and reporting susceptibility values. The magnitude and phase images shown are the same as those in Figure 4.

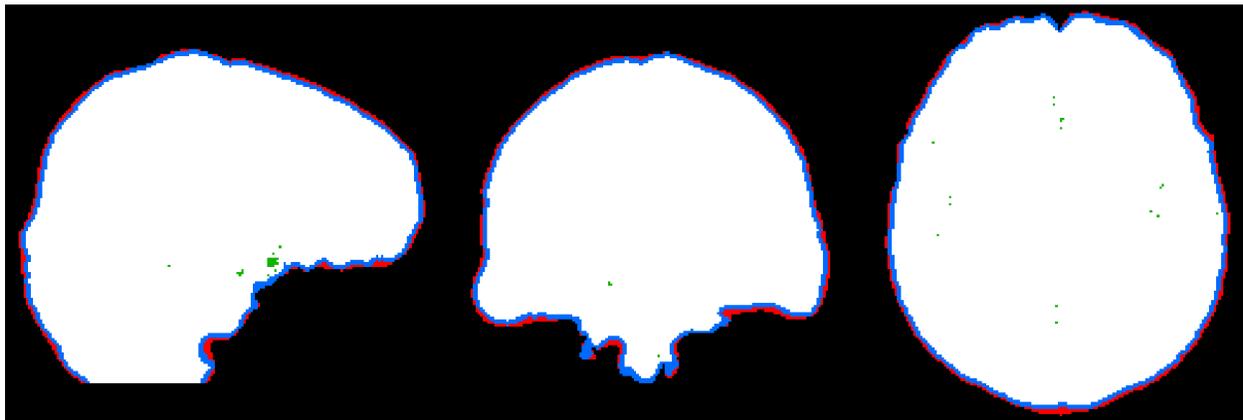

**Figure 6:** By using the mask or reliable phase (Mask 2), the initial BET mask (Mask 1) can be further improved by removing unreliable phase data near the boundary (red). Mask 3 is used for background field removal (BFR). After BFR, Mask 3 may need to be further eroded depending on the output of the background field removal algorithms (eroded region shown in blue). This is used for visualization of the results and reporting. Unreliable phase data inside the brain can also be masked out for dipole inversion (holes in green, with the final Mask 4 used for dipole inversion in white).

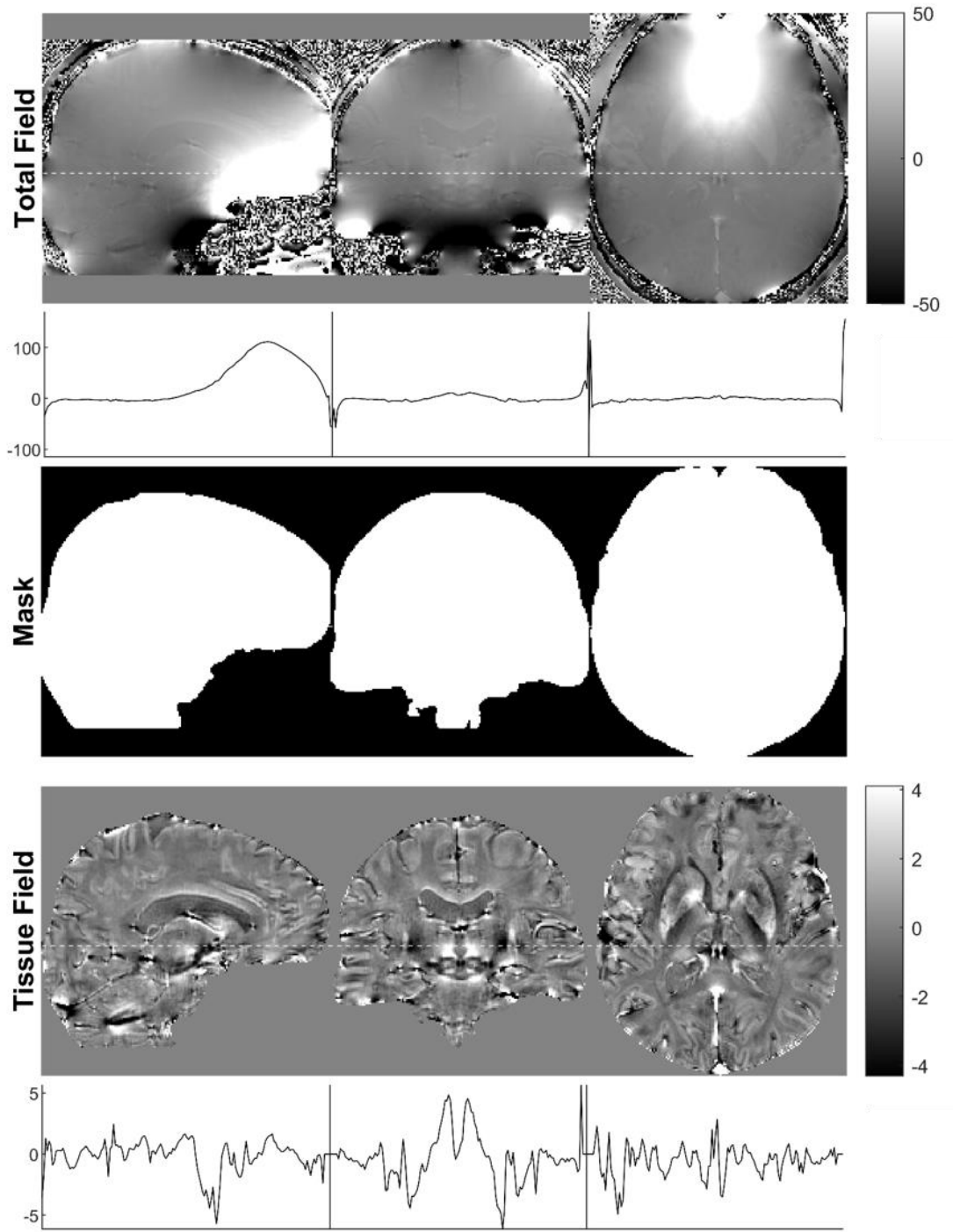

**Figure 7.** Process of background field removal estimates the background field component of the total field (first row; same images as shown on the right-hand side of Figure 4; unit is Hz) relative to a chosen region of interest (brain mask, third row) and subtracts it from the total field, resulting in the tissue field (fourth row; unit is Hz). The tissue field encodes the spatially varying susceptibility within the brain but is much smaller than the background field. This is illustrated by showing cross-sections (indicated by the dotted lines in the field images) in the total field (second row) and the tissue field (last row). The background field was calculated using the V-SHARP method.

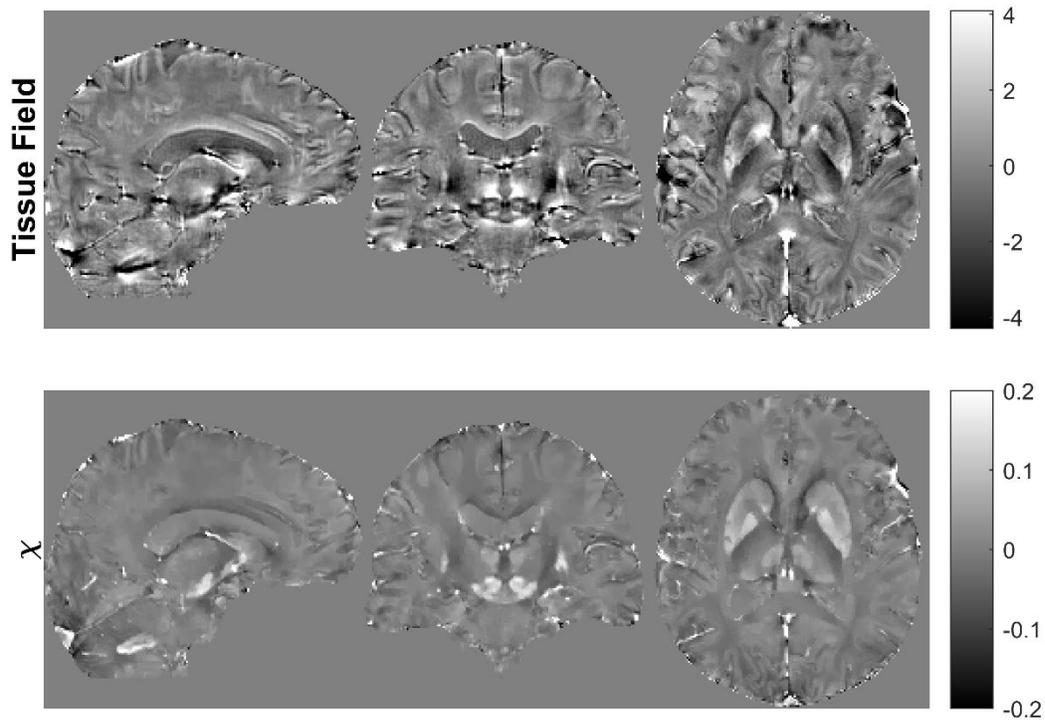

**Figure 8.** The process of dipole field inversion starts from the tissue field (first row, same images as shown in the bottom row of Figure 7; unit is Hz) and estimates the susceptibility map (second row; unit is ppm).

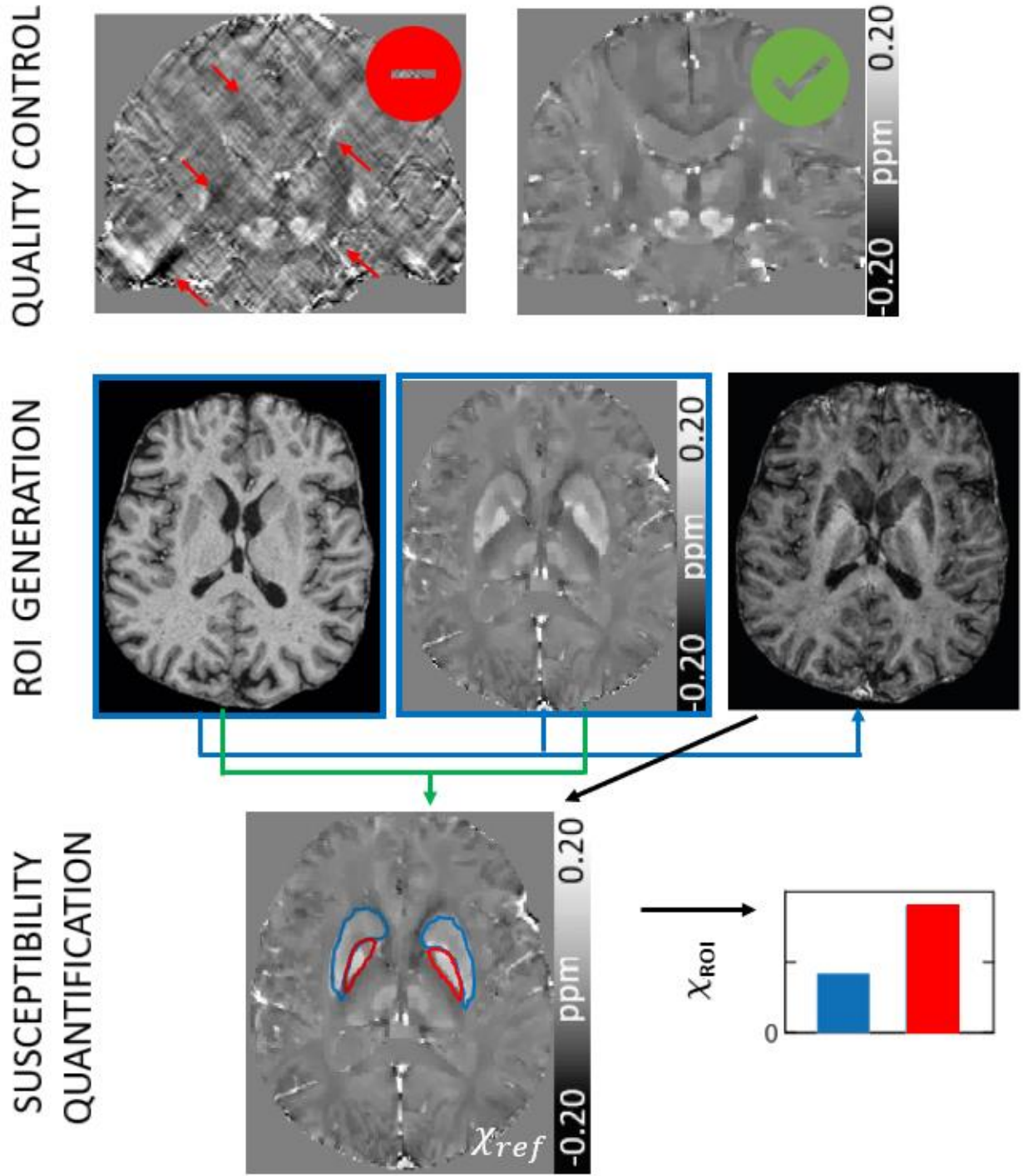

**Figure 9:** Schematic for susceptibility map analysis in case of a study interested in susceptibility values of the putamen (blue ROI on the susceptibility map, $\chi_{ref}$) and globus pallidus (red ROI). Data with streaking artifacts that affect the ROIs need to be excluded (or recalculated when applicable to all study data). ROI generation benefits from the inclusion of susceptibility contrast, e.g., by calculation of hybrid images (blue) or use of $T_1$-weighted and susceptibility data (green). Susceptibility maps need to be referenced, then regional average susceptibility values ($\chi_{ROI}$)

can be computed from referenced susceptibility maps ($\chi_{ref}$). The shown susceptibility map without artifacts is the same as the one in Figure 8.

# Tables

**Table 1.** Reconstruction artifacts, possible sources, and strategies to identify, mitigate these artifacts and criteria to exclude the data.

| Artifact | Streaking and shadowing artifacts 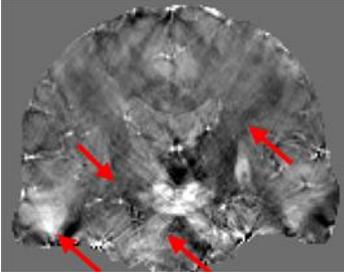 | Incorrect susceptibility values 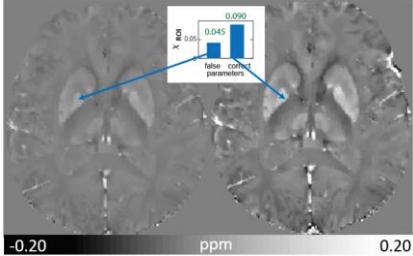 | (Regional) strong noise 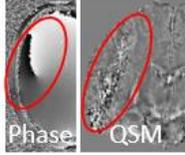 |
|---|---|---|---|
| **Typical sources** | - large susceptibility differences (air-tissue, calcification, hemorrhage etc.)<br>- poor brain mask<br>- use of late TEs and dynamic field fluctuations (especially at higher field strength) in combination with unsuitable mask<br>- inversion algorithm unsuitable for the data (e.g. presence of strong susceptibility sources)<br>- poor choice of reconstruction parameters (for phase unwrapping, background field removal or inversion algorithm)<br>- incorrect coil combination | - Mismatch between acquired and assumed TE values during the reconstruction, e.g., due to unanticipated acquisition protocol changes | - incorrect coil combination<br>- suboptimal image acquisition (coverage, 2D) |
| **Identification** | - manual/visual quality control<br>- automatic detection of outlier regions on phase images or QSM [220] (beware of outliers due to pathology)<br>- automated histogram analysis<br>- use of image quality measures such as the structural similarity index<br>- manual/visual quality control | - outlier detection (based on mean/median ROI values) | (see streaking and shadowing artifacts) |
| **Mitigation** | - adjust masking, reconstruction algorithm and parameters (e.g., exclude late TEs from echo combination)<br>- use appropriate coil combination | - verification of imaging parameters from DICOM header (manually or automatic)<br>- pull imaging parameters from data instead of hard coding it in the pipeline | - use appropriate coil combination<br>- use of recommended acquisition protocol |
| **Data exclusion** | - exclude if ROI affected and recalculation of the entire study cohort with the adjusted pipeline not possible (to avoid bias) | | |

**Table 2.** Commonly used reference regions in the literature

| Reference region | advantages | disadvantages |
|---|---|---|
| | | |
| cerebrospinal fluid [193,205,230,248–252] | · automatic pipelines available [193]<br>· no orientation dependence<br>· susceptibility of CSF unlikely to be significantly affected by disease | · ventricles can be small in young subjects, resulting in segmentation inaccuracies<br>· partial volume effect because of possibly small ventricles in young subjects or compression of ventricles by pathology<br>· CSF flow artifacts<br>· choroid plexus can affect CSF susceptibility assessment in lateral ventricles |
| global white matter regions (not restricted to internal capsule) [28,91,253] | · large region | · orientation dependence<br>· might be affected by pathology, e.g. demyelination, gliosis, hemorrhage, atrophy |
| internal capsule [40,254–256] | | · orientation dependence<br>· might be affected by pathology, e.g. demyelination, gliosis, hemorrhage, atrophy, focal lesions<br>· Relatively small region |
| whole brain [229,236,236,257] | · no extra mask required, brain mask from previous processing steps can be used<br>· intrinsic for some methods<br>· large region | · might be affected by pathology and age (e.g. myelination, global demyelination, gliosis, iron accumulation, hemorrhage)<br>· due to large WM fraction similar limitations as "white matter" above. |

**Table 3.** Recommendations for reporting of parameters of the acquisition hardware.

| Item | Notes and examples | Recommended |
|---|---|---|
| Field strength | DICOM tag (0018,0087) | essential |
| Vendor | DICOM tag (0008, 0070) | essential |
| Scanner model | DICOM tag (0008, 1090) | 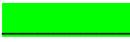 |
| Software release | DICOM tag (0018, 1020) | 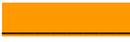 |

| Type of coil(s) used, including information on number of channels | e.g. "... a transmitting body-coil and a 64-channel head-and-neck receiving coil" | <span style="background-color:#00ff00">     </span> |
|---|---|---|
| Gradient system | e.g. "... a gradient system with maximum amplitude = 50 mT/m and slew rate = 200 mT/m/ms" | <span style="background-color:#ff0000">     </span> |

**Table 4.** Recommendations for reporting of parameters of the acquisition sequence.

| Item | Notes | Recommended |
|---|---|---|
| Acquisition sequence type | 2D *vs* 3D; GRE *vs* EPI etc.; | essential |
| Acquisition sequence commercial name | e.g. "SWAN", "MERGE", "SWIp"… | <span style="background-color:#00ff00">     </span> |
| k-space sampling trajectory scheme | cartesian *vs* spiral *vs* radial etc. | essential, if not cartesian |
| Acquisition orientation | pure axial *vs* sagittal *vs* oblique | <span style="background-color:#00ff00">     </span> |
| Number of echoes, $TE_1:\Delta TE:TE_{max}$ | e.g. 7 echoes, TE = 5:5:35 ms | essential |
| TR | | essential |
| FA | | <span style="background-color:#00ff00">     </span> |
| Pixel Bandwidth or Receiver Bandwidth [Hz] | DICOM tag (0018, 0095) | <span style="background-color:#00ff00">     </span> |
| Spatial coverage (FOV) and acquisition matrix size | | essential |
| Voxel Size | Attention: it can be different from "FOV divided by matrix size" | essential |
| Monopolar *vs* bipolar echoes | Indicate if the sequence produces monopolar of bipolar echoes | <span style="background-color:#00ff00">     </span> |
| Average ± std center frequency [MHz] | In multi-scanner studies, mean ± std center frequency shall be reported for data from each scanner. For example, Siemens "3T" scanners systematically operate at <2.9T. DICOM tag (0018,0084) | essential for multi-scanner studies; unnecessary for single-scanner studies |
| Flow compensation | Yes / no; if yes, please indicate the compensated echo(s): all *vs* only the first one; and direction (full, phase) | <span style="background-color:#00ff00">     </span> |

| Acceleration type and factor | Yes / no. If yes: SENSE (or ASSET) *vs* GRAPPA (or ARC), compressed SENSE, etc.; indicate phase factor and slice factor (if 3D) | essential |
|---|---|---|
| Partial Fourier factor | Use should be avoided. If used, indicate partial Fourier factors in phase and slice direction | essential, if used |
| Partial echo (GE/Philips) *aka* Asymmetric echo (Siemens) *aka* Half echo (Hitachi) | Use should be avoided. | essential, if used |
| Elliptical k-space shutter | Yes / no. | 🟩 |
| Phase stabilization | Option available only in particular implementations. If the option is available, indicate Yes / no | 🟧 (essential if used) |
| Excitation pulse | Fat-sat *vs* Water-only | 🟧 |
| Scan duration |  | essential |

**Table 5.** Recommendations for reporting of parameters of the reconstruction and analysis pipelines.

| Item | Notes | Recommended |
|---|---|---|
| Toolbox used | Specify toolbox name and version (or download date), e.g. FANSI, STISuite, MEDI, etc. | mandatory |
| Algorithms used | For each step of the recon pipeline (phase reconstruction, echo combination, masking, phase unwrapping, background field removal, dipole inversion), please specify the algorithm used. Indicate the numerical values of relevant parameters (even if default values were used), e.g. regularization parameters. | mandatory, at least for non-default algorithms and parameters |
| Further processing | If further processing was necessary to make images compatible with image review environments (such as PACS) used in the study, any data manipulation (including geometrical transformations, interpolation, header data changes, etc.) should be reported | 🟥 |
| Referencing | Magnetic susceptibility values should always be reported in either ppm or ppb (parts-per-billion) and the reference region (see the Section 8) should be explicitly stated, even in the case the adopted method did implicit whole brain referencing.<br><br>When the reference region used in the study is not the whole-brain mask, its [mean ± std] susceptibility value when referenced to the whole-brain mask should be reported, to enable post-hoc re-referencing for meta-analyses. Generally, it should be discussed in the Discussion section how potential pathological changes within the reference region may have biased the study outcome. | mandatory |
| Data inclusion/exclusion criteria | Details on data inclusion/exclusion criteria should be reported. For example: which artifacts were taken into consideration, and which level of artifact severity was considered as a threshold for inclusion/exclusion. The description of this aspect, which is study-specific, can be supported by images with representative cases in the Supplementary Materials. | Mandatory, in studies where datasets were excluded based on image quality, or when datasets with visible artifacts were deemed acceptable for inclusion |

# Supplementary Materials I

## S1.1 Detailed Consensus Committee Author Contributions

This section lists the core authors of each section along with the section's lead author(s). Core authors are the members of the QSM Consensus Organization Committee who developed the initial draft of the consensus recommendations and incorporated the feedback from the QSM community (see Acknowledgement section). Lead authors orchestrated the creation of the section and are responsible for the final version. Authors are listed in alphabetical order.

**Overall facilitation, organization, and coordination:** Ferdinand Schweser, Yi Wang

**Section "Pulse Sequences and Protocol Recommendation":**
**Lead authors**: Jongho Lee, Jose Marques.
**Core authors:** Berkin Bilgic, Mauro Costagli, Christian Langkammer, Chunlei Liu, Simon Robinson, Ferdinand Schweser, Karin Shmueli, Pascal Spincemaille, Sina Straub

**Section "Coil Combination, Saving and Exporting":**
**Lead author**: Simon Robinson
**Core authors:** Mauro Costagli, Ferdinand Schweser, Karin Shmueli, Pascal Spincemaille

**Section "Phase Unwrapping and Echo Combination":**
**Lead author:** Xu Li
**Core authors:** Christian Langkammer, Chunlei Liu, Carlos Milovic, Simon Robinson, Ferdinand Schweser, Karin Shmueli

**Section "Creation of Masks":**
**Lead author:** Carlos Milovic
**Core authors:** Christian Langkammer, Simon Robinson, Karin Shmueli

**Section "Background field removal":**
**Lead author:** Pascal Spincemaille

**Core authors:** Carlos Milovic, Ferdinand Schweser

**Section "Dipole Inversion":**
**Lead author:** Yi Wang
**Core authors:** Xu Li, Carlos Milovic, Pascal Spincemaille

**Section "Analysis of susceptibility maps":**
**Lead author:** Sina Straub
**Core authors:** Mauro Costagli, Christian Langkammer, Xu Li, Ferdinand Schweser

**Section "Presentation and Publication":**
**Lead author:** Mauro Costagli
**Core authors:** Jeff Duyn, Christian Langkammer, Xu Li, Chunlei Liu, Ferdinand Schweser, Sina Straub

**Code implementation (Supplementary Materials II):**
**Lead author:** José Marques
**Core authors:** Kwok-Shing Chan

## S1.2 Author Conflict of Interest Statements

In the interest of transparency, this section provides disclosures of perceived or actual risks of a conflict of interest of the authors.

**Berkin Bilgic** is the first author of the Wave-CAIPI publication (Bilgic et al., 2017) mentioned in Section 2.3 among the methods that provide drastic decreases in acquisition time.

**Mauro Costagli** has no conflicts of interest to declare.

**Christian Langkammer** has no conflicts of interest to declare.

**Jongho Lee** is the corresponding author of the papers for QSMnet, which is a deep learning-powered QSM reconstruction method (Yoon et al., 2018), and χ-separation, which is an advanced susceptibility mapping method for susceptibility source separation (Shin et al., 2021).

**Xu Li** is the first author of the paper on using both T1-weighted and susceptibility contrast multi-atlas in image analysis (Li et al., 2019), recommended in Section 8.

**Chunlei Liu** co-authored papers on VSHARP, Laplacian-based phase unwrapping and multi-echo weighted phase combination methods. He is a co-author of the STI Suite software and co-inventor of QSM-related patents.

**José Marques** is the senior author of the publication associated with the software (SEPIA) used in the Code Implementation section (Chan and Marques, 2021), and first author of the in-silico frame work to evaluate QSM reconstruction pipelines used in the QSM Challenge 2.0 (Marques et al., 2021).

**Carlos Milovic** is the author and manager of the FANSI toolbox, which is recommended in several sections. This includes the FANSI (Milovic et al., 2018) and Weak Harmonics (Milovic et al., 2019) algorithms, for which he is the first author.

**Simon Robinson** is the senior author of the coil combination methods MCPC-3D-S and ASPIRE (Eckstein et al., 2018) (Eckstein et al., 2019) which are mentioned in Section 2 as alternatives to the recommended "prescan normalize and adaptive combined". He is also the senior author of the phase unwrapping method ROMEO (Dymerska et al., 2021), which is described in Section 4 and which incorporates the steps in the "weighted echo averaging with template unwrapping" approach; one of the two recommended approaches to echo combination. He is also a co-author on two papers proposing the use of phase-based quality metrics in masking (Hagberg et al., 2022; Stewart et al., 2022) in Section 5.

**Ferdinand Schweser** is the first author of the original publication introducing the SHARP technique (Schweser et al., 2011). The SHARP technique is the basis of the VSHARP technique recommended in Section 6. He is the first author of a paper that introduces an L1-norm type regularization dipole inversion algorithm (Schweser et al., 2012) recommended in Section 7. He is the last author of papers on using both T1-weighted and susceptibility contrast in image analysis (Feng et al., 2017; Hanspach et al., 2017), recommended in Section 8. He is the last author of the in-silico frame work to evaluate QSM reconstruction pipelines used in the QSM Challenge 2.0 (Marques et al., 2021). He has research support from Philips Medical Systems Nederland B.V.

**Kwok-Shing Chan** is the first author of the publication associated with the software (SEPIA) used in the Code Implementation section (Chan and Marques, 2021).

**Karin Shmueli** is an author of the phase unwrapping method ROMEO (Dymerska et al., 2021), which is described in Section 4 and which incorporates the steps in the "weighted echo averaging with template unwrapping" approach; one of the two recommended approaches to echo combination.

**Pascal Spincemaille** is co-author on the publications describing the PDF, LBV and MEDI methods. He is co-inventor on QSM-related patents owned by Cornell University. He is consultant for and has ownership share in MedImageMetric LLC.

**Sina Straub** has no conflicts of interest to declare.

**Peter van Zijl** has research support from Philips and technology licensed to Philips.

**Yi Wang** is co-author on the publications describing the PDF, LBV and MEDI methods. He is co-inventor on QSM-related patents owned by Cornell University. He is consultant for and has ownership share in MedImageMetric LLC.

## S1.3 Approach and History of the Consensus Paper

The idea to create a community-driven recommendations paper for the implementation of QSM gained traction through email conversations on the email distribution list of the program committee for the 2022 Joint Workshop on MR phase, magnetic susceptibility and electrical properties mapping. One of the members of the later established QSM Consensus Organization Committee (YW) revived the idea to create a consensus or white paper for QSM, which had previously been proposed during the standardization session at the 2019 International Workshop on MRI Phase Contrast and Quantitative Susceptibility Mapping held in Seoul, Korea. The discussion quickly separated from the program committee, with the group involved growing to include 13 interested volunteers. The first meeting for the project was organized by two members of the QSM Consensus Organization Committee (FS and YW) and occurred virtually on March 24, 2022. At this first meeting, the group agreed on the scope, title, the main paper sections, an action plan for achieving consensus and writing the paper, and a timeline. After the meeting, all participants specified which of nine paper sections they would be interested in developing for the QSM community, and whether they would consider a leading role in this process. Each section team was tasked with developing a first set of recommendation statements to be disseminated to the ISMRM Electro-Magnetic Tissue Properties Study Group as a starting point for achieving consensus. Eight sections for the paper were identified that corresponded to the segments of the QSM acquisition and processing pipeline: 1) Pulse Sequences and Protocol Recommendation, 2) Coil Combination, Saving + Exporting, 3) Phase Unwrapping and Echo Combination, 4) Masking, 5) Background field removal, 6) Inversion, 7) Analysis of susceptibility map, and 8) Presentation and Publication. Lead authors and contributing core authors for each section are listed in the Supplementary Materials Section S1.1 above.

The QSM Consensus Organization Committee reconvened on April 20, 2022 and agreed to disseminate the consensus recommendations to the QSM community for feedback at the business meeting of the ISMRM Electromagnetic Tissue Properties Study Group on June 3, 2022. The group agreed to revise the consensus recommendations and wrote, for each section, a brief overview section of the subject matter, providing a first full draft of the consensus recommendations. At the ISMRM 2022 annual meeting, a subset of the committee met in person to discuss the consensus paper project in person (May 8, 2022), and various members attended the session on white papers at the conference to better understand the requirements of such a format both regarding its preparation and the final product.

At the study group meeting on June 3, 2022, the consensus initiative was presented (FS) and a draft of the consensus recommendations was made available to the study group members as an editable online document. Study group members and non-members of all career levels with expertise in QSM were invited to contribute to the manuscript by reviewing it and suggesting modifications or extensions for the QSM Consensus Organization Committee to incorporate. The link to the online document was provided at the study group meeting and distributed after the study group meeting through the study group mailing list. After the open feedback period, the QSM Consensus Organization Committee incorporated the feedback received, discussed open questions, and, on 7/14/2022, defined the final timeline aiming for submission of the manuscript shortly after the 2022 Joint Workshop on MR Phase, Magnetic Susceptibility and Electrical Properties Mapping in Lucca, Italy (10/16-10/19/2022). The committee revised the manuscript through several review iterations in which either only section leaders or all core-authors were involved. Some section leaders chose to perform surveys among the committee members to quantify and resolve disagreement on controversial statements or requests from the community for which the section sub-groups could not arrive at a unanimous recommendation. All surveys and their outcomes were reported in the manuscript.

The first complete version of the manuscript was provided to all participants of the 2022 Joint Workshop on MR Phase, Magnetic Susceptibility and Electrical Properties Mapping via email on 10/10/2022. The manuscript was discussed in personal communications at the workshop and a summary of the recommendations was presented in Session 9 of the workshop (10/19) followed by an open discussion with the workshop attendees. The committee incorporated the feedback collected from workshop attendees and distributed the resulting manuscript to all members of the EMTP SG via the study group mailing list on 11/23/2022 with the request to disseminate the manuscript further to interested parties and to provide feedback via email by 12/15/2022. During this period, the committee actively reached out to the industry to confirm vendor-specific statements in the manuscript and seek additional input from industry representatives on the manuscript. On 12/16, the committee held an EMTP SG Virtual Meeting which presented the final manuscript, including background information on the process employed by the committee, as well as a summary of all consensus recommendations. Between 12/16/2023 and 3/30/2023, the committee incorporated all suggestions from the October/November feedback period and finalized the manuscript for submission and official endorsement by the ISMRM EMTP study group. In addition to the study group endorsement, the committee distributed an online form to provide the opportunity of endorsement to EMTP SG non-members, listed in Section S1.4 below.

# S1.4 Individuals Endorsing the Consensus Who Are Not Members of the ISMRM EMTP Study Group

In the published paper, this section will include a list of all non-members who endorsed the manuscript after the study group endorsement.

## S1.5 Suggested Protocols

Please note that the protocols below are simply suggestions (with timings rounded to the nearest millisecond) and that the parameters can, of course, be adjusted where needed, according to the principles in Section 2, to accommodate differences in scanner hardware and software.

### 1.5T Protocol

TR = 50 ms; $TE_1$ = 5 ms; Echo spacing = 10 ms; Number of echoes = 5; flip angle = 23°

Matrix size (AP LR HF) = 176 x 140 x 114

Resolution = isotropic 1.2 mm

Parallel imaging factor = 2; elliptical k-space shutter

Acquisition time = 6 mins

### 3T Protocol

TR = 33 ms; $TE_1$ = 5 ms; Echo spacing = 6 ms; Number of echoes = 5; flip angle = 15°

Matrix size (AP LR HF) = 256 x 176 x 144

Resolution = isotropic 1 mm

Parallel imaging factor = 2; elliptical k-space shutter

Acquisition time = 6 mins

### 7T Protocol

TR = 25 ms; $TE_1$ = 4 ms; Echo spacing = 4 ms; Number of echoes = 5; flip angle = 10°

Matrix size (AP LR HF) = 296 x 234 x 190

Resolution = isotropic 0.75 mm

Parallel imaging factor = 3; elliptical k-space shutter

Acquisition time = 5 mins

## S1.6 Resources for Susceptibility Map Analysis

### Structural/ROI segmentation

- Uses QSM contrast:

    - Advanced Normalization Tools - ANTs (https://github.com/ANTsX/ANTs)
    - SuscEptibility mapping PIpeline tool for phAse images – SEPIA (https://github.com/kschan0214/sepia) (uses ANTs for atlas-based segmentation)
    - https://mricloud.org/
    - STI Suite (https://people.eecs.berkeley.edu/~chunlei.liu/software.html)

- Not QSM-specific/-based, uses T1-weighted images:

    - Fastsurfer (https://github.com/Deep-MI/FastSurfer)
    - Neurodesk (https://github.com/NeuroDesk) (ROI segmentation is based on T1-weighted images, although it includes ANTs)
    - QSMxT (https://github.com/QSMxT/QSMxT) (ROI segmentation is based on T1-weighted images, although it includes ANTs)
    - Freesurfer (https://surfer.nmr.mgh.harvard.edu/)
    - FSL (https://fsl.fmrib.ox.ac.uk/fsl/fslwiki)
    - SPM12 (http://www.fil.ion.ucl.ac.uk/spm/software/spm12/)
    - A Computational Anatomy Toolbox for SPM – CAT (https://neuro-jena.github.io/cat/)

### Lesion segmentation

- Uses QSM contrast:

    - QSMRim-Net (https://github.com/tinymilky/QSMRim-Net)

- Not QSM-specific/-based, uses T1-weighted images:

    - LST - A lesion segmentation tool for SPM (https://www.applied-statistics.de/lst.html)
    - NicMSlesions (https://github.com/sergivalverde/nicMSlesions/)
    - Freesurfer (https://surfer.nmr.mgh.harvard.edu/fswiki/Samseg)

### QSM-based vessel segmentation

- Uses QSM contrast:

    - CVI-MRI (https://github.com/philgd/CVI-MRI)

- Nighres (https://github.com/nighres/nighres)
- GRE_vessel_seg (https://github.com/SinaStraub/GRE_vessel_seg)

### Voxel based analysis

- Uses QSM contrast:

- Advanced Normalization Tools - ANTs (https://github.com/ANTsX/ANTs)

- Not QSM-specific/-based, uses T1-weighted images:

- Freesurfer (https://surfer.nmr.mgh.harvard.edu/)
- FSL (https://fsl.fmrib.ox.ac.uk/fsl/fslwiki)
- SPM12 (http://www.fil.ion.ucl.ac.uk/spm/software/spm12/)
- A Computational Anatomy Toolbox for SPM – CAT (https://neuro-jena.github.io/cat/)

<> </>

# Supplementary Materials II – Example Data and Code for QSM Reconstruction

## Introduction

These supplementary materials provide an overview of the QSM reconstruction processing, from scanner-provided data to QSM maps, based on the recommendations present in the main text. This document has two main purposes:

(1) Allowing readers to reproduce the results shown throughout the paper, and

(2) Providing readers with the means to reconstruct their own data using the recommended processing with data acquired on any of the 3 major MR providers broadly following the recommendations.

Full datasets, results and processing scripts are available on Zenodo:
https://doi.org/10.5281/zenodo.7410455
Example data of version v0.2.1 were used in this paper.

## S2.1 Data availability

Data are available from scanners of three vendors: GE, SIEMENS, and PHILIPS, acquired with the recommended protocol described in Section 2 and Supplementary Materials I, Section S1.5. For each vendor, both monopolar and bipolar readout strategies were used to acquire the data for demonstration purposes. The data from GE and SIEMENS scanners were not pre-scan normalized (which does not follow the recommendations), while the PHILIPS data have two normalization methods applied. In this way, we demonstrate the robustness of the proposed pipeline to a variety of implementations of the recommended protocol.

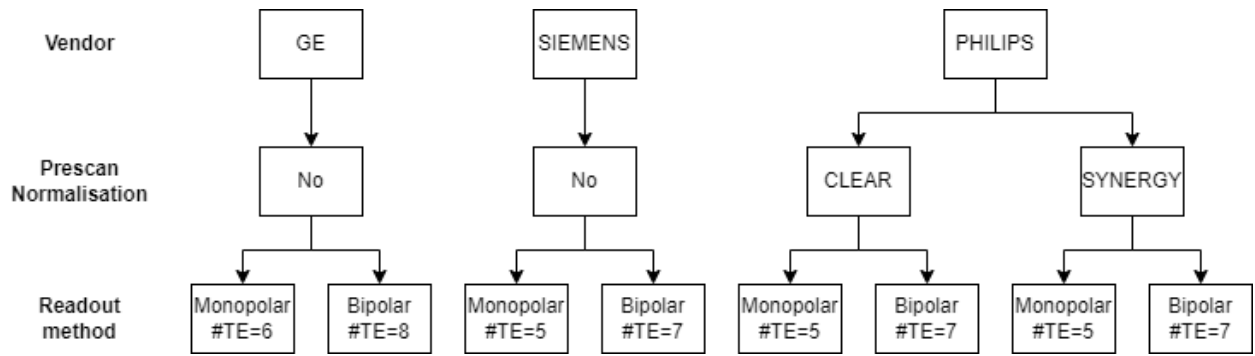

Figure S2.1: An illustration of the raw data available.

## S2.2 Data preparation and organization

Background

There are two zip-files available in Zenodo containing the example data organised in two different ways:

(a) "QSM_CONSENSUS_Paper_Example_DICOM_code.zip"

This zip file contains example DICOM images exported from the scanners without any modifications. The code directory accompanied with this file contains the scripts to (1) convert the DICOM images to NIFTI format, (2) organise the NIFTI images according to BIDS v1.8.0, and (3) perform QSM reconstruction.

(b) "QSM_Consensus_Paper_example_Data_Result_Code.zip"

This zip file contains all data and results that were produced by running all the scripts provided in the code directory.

The following Data Preparation section provides information on all the pre-processing steps to prepare the unmodified DICOM images to the BIDS format data that is ready for QSM reconstruction in SEPIA. If the readers are interested in the QSM reconstruction and work on the "QSM_Consensus_Paper_example_Data_Result_Code.zip" file only, they may skip Section "Data Preparation".

## Data Preparation

The scripts created for this section were tested on a Mac system (macOS 13.2) and a Linux system (CentOS 7). They were not tested on Windows systems and would require adaptations to work there.

Most imaging software in the field typically deals with images in Analyze or NIfTI format. As such the raw data (imaging data after coil combination) provided has to be converted to this format in 4 steps:

**Step 1: Unzip the received data and reformat the directory structure**
**Script**: Preparation_01_rename_received_data.sh

**Step 2: Convert DICOM images into NIfTI format**
**Dependency**: dcm2niix (version 1.0.20220720)
**Script**: Preparation_02_convert_dicom2nii.sh

**Step 3: Rename the files according to the BIDS format** (Brain Imaging Data structure)
**Dependency**: Matlab R2016b onwards
**The naming strategy is as follows**:
- Vendors are identified using the session tag: **ses-<GE|PHILIPS|SIEMENS>**
- For **GE** and **SIEMENS**, different readout methods are identified using the acquisition tag: **acq-<Bipolar|Monopolar>**;
- For **PHILIPS**, the normalisation method is also printed on the acquisition tag, i.e., **acq-<BipolarCLEAR|BipolarSYNERGY|MonopolarCLEAR|MonopolarSYNERGY>**

**Script**: Preparation_03_rename_to_bids_format.m

**Step 4: Prepare NIFTI data for SEPIA**

**Dependency**: (1) Matlab R2016b onwards, (2) SEPIA v1.2.2.4

**Involves the following operation**:

- Combining individual multi-echo 3D volumes into a single 4D volume with TE in the 4th dimension;
- Obtaining header info (e.g., $B_0$ direction and TE) from NIfTI header and JSON sidecar files and saving as SEPIA's header format;
- (GE only) Correcting inter-slice opposite polarity on real and imaginary images and exporting phase images from the corrected real/imaginary data

**Script**: Preparation_04_prepare_for_sepia.m

## Data organization

The following tree diagram illustrates the directory structure of how the data are organised after running all the scripts provided in the code directory "QSM_Consensus_Paper_Example_Code/". The content of the different directories is mentioned after the comment "%" symbol. Note that similar directories exist under the "/derivatives/SEPIA/SIEMENS/" and "/derivatives/SEPIA/PHILIPS/" as under "/derivatives/SEPIA/GE/".

```
QSM_Consensus_Paper_Example_DICOM_Code/
|-- QSM_CONSENSUS_DATA.zip             % Zip file containing all unmodified DICOM images
|-- protocols                          % Protocol text/HTML files
|-- QSM_Consensus_Paper_Example_Code   % Containing all the scripts
|   |-- doc                            % Containing manual to use the Example data
|   |-- From_DICOM_zip_file_to_SEPIA_ready  % Scripts for preparing QSM_CONSENSUS_DATA.zip
|   |-- SEPIA_Pipeline_FANSI           % SEPIA pipeline config files with FANSI recon
|   `-- SEPIA_Pipeline_MEDI            % SEPIA pipeline config files with MEDI recon
|-- raw                                % DICOM images
|-- converted                          % dcm2niix output
|   |-- GE
|   |   |-- Bipolar                    % Bipolar readout acquisition
|   |   `-- Monopolar                  % Monopolar readout acquisition
|   |-- PHILIPS
|   |   |-- Bipolar_CLEAR              % with CLEAR normalisation
|   |   |-- Bipolar_SYNERGY            % with SYNERGY normalisation
|   |   |-- Monopolar_CLEAR
```

```
|   |   `-- Monopolar_SYNERGY
|   `-- SIEMENS
|       |-- Bipolar
|       `-- Monopolar
`-- derivatives                        % directory contains all derived output
    `-- SEPIA                          % SEPIA output
        |-- GE
        |   |-- Bipolar
        |   |   `-- GRE
        |   |       |-- Pipeline_FANSI    % Full QSM recon using FANSI for dipole inversion
        |   |       `-- Pipeline_MEDI     % Full QSM recon using MEDI for dipole inversion
        |   `-- Monopolar
        |       `-- GRE
        |           |-- Pipeline_FANSI    % Full QSM recon using FANSI for dipole inversion
        |           `-- Pipeline_MEDI     % Full QSM recon using MEDI for dipole inversion
        |-- PHILIPS
        `-- SIEMENS
```

## 2.3 QSM reconstruction pipeline

This section describes all the QSM reconstruction processing steps performed in SEPIA. All the processing steps are specified in the SEPIA pipeline configuration files, which are in the sub-directories of the script directory: *'QSM_Consensus_Paper_Example_Code/SEPIA_Pipeline_FANSI/'* and *'QSM_Consensus_Paper_Example_Code/SEPIA_Pipeline_MEDI/'*, corresponding to the two processing pipelines demonstrated as follows.

Environment and dependencies

The data were processed using the following set-up:

**Operating system**:
- Linux CentOS 7

**Environment:**
- Matlab R2021a (but the scripts are backwards compatible with earlier Matlab versions from R2016b to R2022a)

**Dependencies:**

The following QSM toolboxes have to be downloaded and integrated into SEPIA following the instruction provided on the SEPIA documentation website (https://sepia-documentation.readthedocs.io/en/latest/getting_started/Installation.html):

- SEPIA v1.2.2.4 (https://github.com/kschan0214/sepia/releases/tag/v1.2.2.4)
- MRITOOLS v3.5.6 (https://github.com/korbinian90/CompileMRI.jl/releases/tag/v3.5.6)
- MEDI toolbox (release: 15th January 2020) (http://pre.weill.cornell.edu/mri/pages/qsm.html)
- FANSI toolbox [v3] (https://gitlab.com/cmilovic/FANSI-toolbox)
- STI Suite v3.0 (https://people.eecs.berkeley.edu/~chunlei.liu/software.html)

QSM reconstruction using Example data

This section describes all the QSM reconstruction settings that were used on the example data. All the methods and algorithm parameters mentioned were already specified in the SEPIA pipeline configuration files (sepia_<GE|PHILIPS|SIEMENS>_<Monopolar|Bipolar>_config.m), which can be found in the sub-directories of the code folder "QSM_Consensus_Paper_Example_Code/": *"SEPIA_Pipeline_FANSI/"* and *"SEPIA_Pipeline_MEDI/"*. Here, we provide an overview of the main parameters of each of these pipelines (Tables S2.1-S2.4) for the readers' convenience.

Processing steps

**Step 1: Preparation**

- (GE only) Phase data must be inverted before QSM reconstruction processing (i.e., phase = -phase), so that paramagnetic susceptibility gives a positive value while diamagnetic susceptibility gives a negative value, same as the data from other vendors. This step was performed with the option provided by SEPIA.
- Brain mask is obtained by using MEDI toolbox implementation of FSL's BET on the 1st echo magnitude image, using default setting -f 0.5 -g 0

- (Bipolar readout data only) Bipolar readout correction based on (Li et al., 2015) using the implementation provided with SEPIA.
- Note that the relevant sequence parameters such as echo time and slice orientation are automatically derived from the data.

**Step 2: Total field estimation and echo combination**

Table S2.1: Algorithm parameters for total field estimation and echo combination.

| Parameters | Values | Remark |
| --- | --- | --- |
| **Echo phase combination** | ROMEO total field calculation | (Dymerska et al., 2020) |
| **MCPC-3D-S phase offset correction** | On | |
| **Mask for unwrapping** | SEPIA mask | FSL's BET mask |
| **Using ROMEO Mask in SEPIA** | Off | |
| **Exclude voxel using relative residual with threshold** | 0.3 (applied on weighting map) | See https://sepia-documentation.readthedocs.io/en/latest/method/weightings.html |

**Step 3: Background field removal**

Table S2.2: Algorithm parameters for background field removal.

| Parameters | Values | Remark |
| --- | --- | --- |
| **Method** | VSHARP | (Li et al., 2011); SEPIA's implementation |
| **Maximum spherical mean value filtering size** | 12 | Unit: voxels |
| **Minimum spherical mean value filtering size** | 1 | Unit: voxels |
| **Remove residual B1 field** | No | |
| **Erode brain mask before BFR** | 1 | Unit: voxel |
| **Erode brain mask after BFR** | 0 | |

## Step 4: Dipole inversion

We demonstrate the dipole inversion steps with two recommended methods (FANSI and MEDI).

### Step 4.1: FANSI dipole inversion

Table S2.3: Algorithm parameters for dipole field inversion using 'SEPIA_PIPELINE_FANSI' pipeline.

| Parameters | Values | Remark |
|---|---|---|
| **Method** | FANSI | (Milovic et al., 2019, 2018) |
| **Iteration tolerance** | 0.1 | |
| **Maximum number of iterations** | 400 | |
| **Gradient L1 penalty, regularisation weight** | 0.0005 | |
| **Gradient consistency weight** | 0.05 | |
| **Fidelity consistency weight** | 1 | |
| **Solver** | Non-linear | |
| **Constraint** | TV | |
| **Method for regularisation spatially variable weight** | Vector field | |
| **Using weak harmonic regularisation** | On | |
| **Harmonic constraint weight** | 150 | |
| **Harmonic consistency weight** | 3 | |
| **Reference tissue** | Brain mask | |

**Step 4.2: MEDI dipole inversion**

Table S2.4: Algorithm parameters for dipole field inversion using 'SEPIA_PIPELINE_MEDI' pipeline.

| Parameters | Values | Remark |
|---|---|---|
| **Method** | MEDI | (Liu et al., 2011) |
| **Regularisation parameter (lambda)** | 2000 | |
| **Method of data weighting** | 1 | SNR weighting |
| **Percentage of voxels considered to be edges** | 90 | |
| **Array size for zero padding** | [0 0 0] | |
| **Performing spherical mean value operator** | On | |
| **Radius of the spherical mean value operation** | 5 | Unit: voxel |
| **Performing modal error reduction through iterative tuning (MERIT)** | On | |
| **Performing automatic zero reference (MEDI+0)** | Off | |
| **Reference tissue** | Brain mask | |

Adaptation of the example pipeline to other studies

The provided SEPIA pipeline configuration file (sepia_<GE|PHILIPS|SIEMENS>_<Monopolar|Bipolar>_config.m) can be reused for

other studies, assuming the data in these studies have the compatible input directory described in the SEPIA documentation website ([https://sepia-documentation.readthedocs.io/en/latest/getting_started/Data-preparation.html](https://sepia-documentation.readthedocs.io/en/latest/getting_started/Data-preparation.html)):

This can be done by updating the "input" variable in the configuration file to the location of the input directory that contains all the essential data in your computer. Alternatively, if a graphical operation is preferred, the SEPIA pipeline configuration files can be imported to the SEPIA's GUI by using the "Load config" button on the bottom left of the GUI display and then select the configuration .m file. The GUI will then be updated to the specified methods and algorithm parameters according to the text in the configuration file. Readers can then specify the required input and output information on the "I/O" panel on the GUI.

## 2.4 Example results

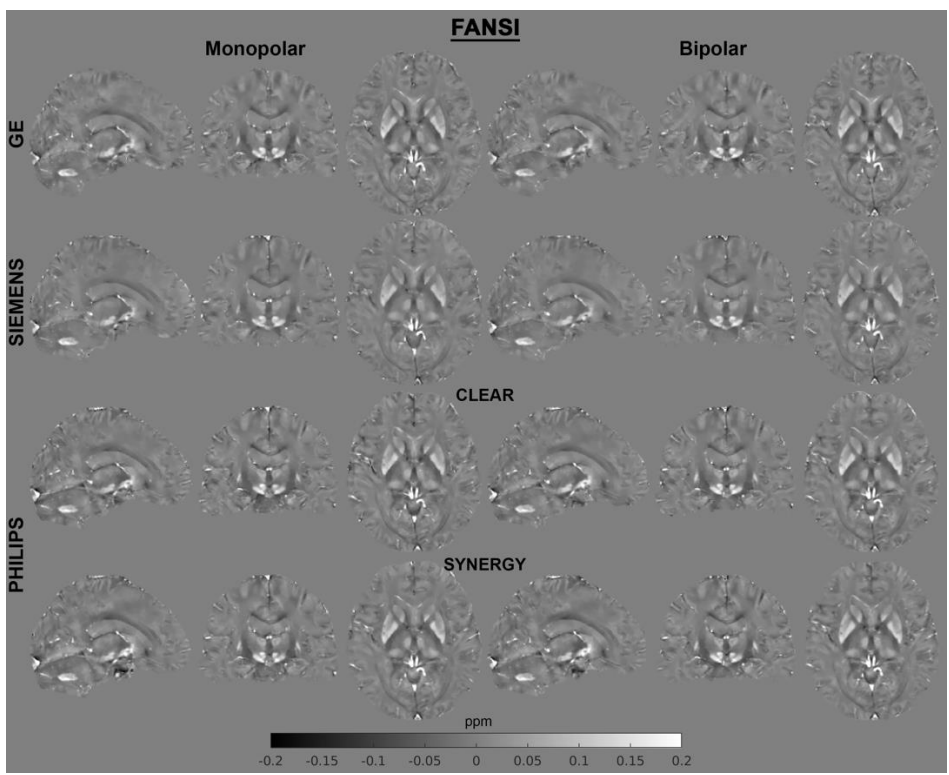

Figure S2.2: Susceptibility maps derived using the "SEPIA_Pipeline_FANSI" processing pipeline.

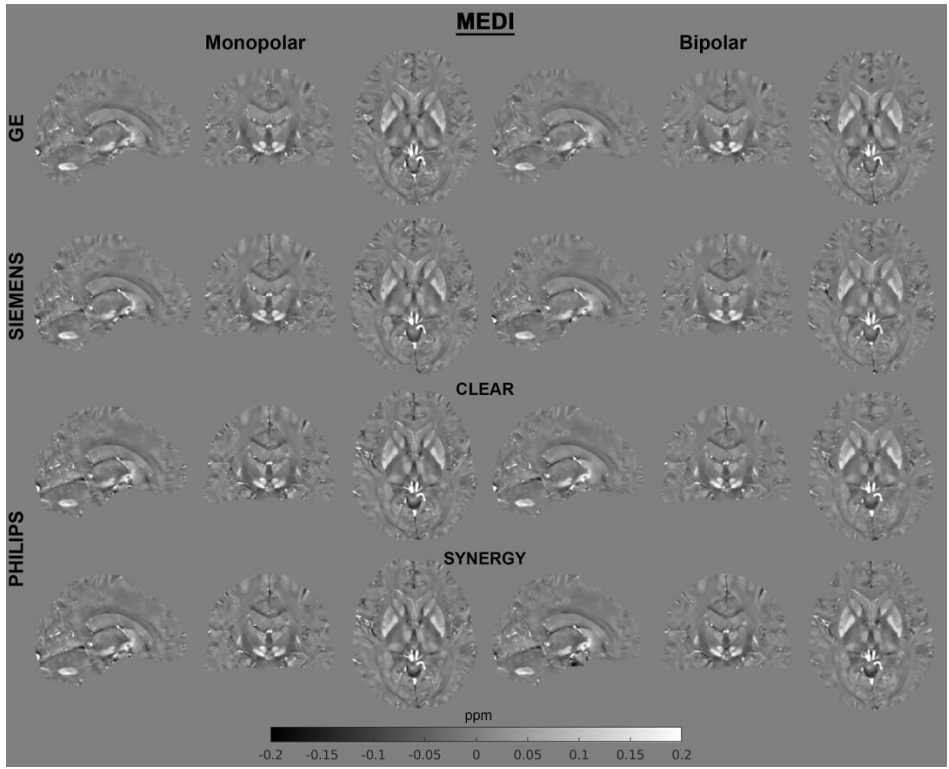

Figure S2.3: Susceptibility maps derived using the "SEPIA_Pipeline_MEDI" processing pipeline.

# Supplementary Materials III - Saving Phase Data and Generating Combined Phase Images: Detailed Information for MR Manufacturers and Systems

## Siemens

**"Adaptive-combined with prescan normalize" (AC-PN)**: This is the recommended approach for systems up to 3T with modern software versions - VE onwards (Jellus V and Kannengiesser S, 2014).

<u>Availability:</u> software version VE11 and later, systems up to 3T.

<u>How to:</u> In the product GRE sequence, i) in the System>Miscellaneous tab, set Coil Combine Mode to Adaptive Combined, ii) in the Resolution>Filter Image tab, check the Prescan Normalize box and iii) in the Contrast>Dynamic tab, set Reconstruction to Magn./Phase.

<u>Limitations:</u> No acceleration is possible in the second phase-encode direction. Not compatible with the SWI option (checkbox on the overview tab).

**"Singular Valued Decomposition Phase Combination"**: An alternative phase combination provided by Siemens for UHF systems, also applicable to single-echo acquisitions. (Inati et al., 2014)

<u>Availability:</u> software version VE11 and later (for research purposes only).

<u>How to:</u> acquire GRE data and retro-reconstruct using the tool TWIX (as advanced user, Windows>Run>twix), changing the ICE program (tICEProgramName) from %SiemensIceProgs%\IceProgram3D to %SiemensIceProgs%\IcePat and set YAPS.AdaptiveCoilCombineAlgo (default -1 -> ACC_ALGO_EVD_PSNPC = 9) to ACC_ALGO_EVD_SVDPC = 5. Does not to work with "meas dependencies" on unless using the "retro recon UI tool" at VE (echo symbol) / NX.

Limitations: None known. For research purposes only. Needs Advanced User privileges.

**"ASPIRE Online"**: A coil combination method for multi-echo data, suitable for systems without a body coil, e.g. UHF (Eckstein et al. 2018).
Availability: VB17, VE11C, VE12U, all field strengths, via C2P (simon.robinson@meduniwien.ac.at).
How to: In the C2P GRE sequence (ke_gre_aspire_*), in the Sequence>Special tab, set Phase Combination to ASPIRE.
Features: Allows acceleration in the second phase-encode direction and provides T2*/R2* mapping.
Limitations: Requires at least two echoes and TE2=2*TE1. For bipolar, requires at least three echoes and TE3=3*TE1, TE2=2*TE1. Needs Advanced User privileges for installation.

**"MCPC-3D-S"/"ASPIRE Offline"**: A coil combination method for multi-echo data, suitable for systems without a body coil, e.g. UHF (Eckstein et al. 2018).
Availability: All systems.
How to: This method needs phase and magnitude data from all channels to be saved and exported for offline processing. In the GRE sequence, i) in the System>Miscellaneous tab, check the Save uncombined box and ii) in the Contrast>Dynamic tab, set Reconstruction to Magn./Phase. Reconstruct phase and magnitude offline using https://github.com/korbinian90/ASPIRE (MATLAB) or the compiled/Julia unwrapping program ROMEO (https://github.com/korbinian90/ROMEO), which will unwrap and also combine data over coils if there is a 5th dimension (x,y,z,echo,coil).
Features: Allows acceleration in the second phase-encode direction, not subject to the echo time constraints of ASPIRE.
Limitations: Requires export of separate channel data (a large number of files).

**"Virtual Reference Coil"**: A coil combination method suitable for systems without a body coil, e.g. UHF (Parker et al. 2014).
Availability: All systems.

How to: This method can be performed online, using a dedicated ICE program which is available via C2P (request to Mathieu Santin; mathieu.santin@icm-institute.org) or offline, in which case phase and magnitude data from all channels need to be saved and exported for offline processing. In the GRE sequence, i) in the System>Miscellaneous tab, check the Save uncombined box and ii) in the Contrast>Dynamic tab, set Reconstruction to Magn./Phase. Reconstruct phase and magnitude offline using https://github.com/mckib2/virtcoilphase.

Features: Allows acceleration in the second phase-encode direction, applicable to single-channel data.

Limitations: Requires export of separate channel data (a large number of files). Can fail in the cerebellum, for large objects or at field strengths above 7T. Needs Advanced User privileges for installation (online version).

**"Multi-echo Coil Combination"**: A coil combination method for multi-echo data, suitable for 3T and 7T systems.

Availability: VB17 via C2P (pas2018@med.cornell.edu).

How to: In the C2P GRE sequence (customer/gre), set Coil Combination to Adaptive Combine.

Features: The sequence produces suitable magnitude and phase DICOM data directly on the scanner. Before channel combination, the phase of the first echo is subtracted from the phase of all echoes after which the channel phases are averaged to obtain the combined phase (Eq 13 in Bernstein et al., 1994). The magnitude is obtained using sum-of-squares.

Limitations: 1) The method only works for 2 or more echoes. 2) The image reconstruction method is memory intensive. Needs Advanced User privileges for installation.

Philips

**"SENSE or CS-SENSE"**: The product 3D FFE sequence allows reconstruction of coil-combined magnitude/phase/real/imaginary data using SENSE or CS-SENSE (compressed-sensing)

Availability: SENSE is available in all systems with V4 and later software versions, CS-SENSE is available in some V5 systems

How to: in the "Postproc" tab -> "Images" -> Select output of "M" and "P" for magnitude and phase output. For conversion with DCM2NII and DCM2NIIX, the "Philips precise scaling" parameter should be set to ON to avoid using the other rescaling factors provided (which only adjust relative pixel intensity but do not provide quantitative rescaled values).

When using the SWIp product sequence (e.g. the clinic, to get SWI images), save the magnitude and phase data using the "Delayed reconstruction" procedure (need to turn "Postproc" tab -> "Save raw data" to "yes"). On Release 5 of the software (R5), this does not require a research key and is performed by right-clicking on the exam card and selecting the delayed recon option.

Limitations: For V5, need to set "Postproc" tab -> "Images" -> "SWIp" to "no" to allow unfiltered phase output for QSM, otherwise have to use the "Delayed reconstruction".

# GE

**"Research sequence"**: The product SPGR sequence allows the reconstruction of mag/real/imag images by enabling the right features in the source code. This requires a research key.

Availability: The Cornell group (pas2018@med.cornell.edu) can share a compiled version with groups with a valid GE research license (RCSL). Software version 14 and newer.

How to: The sequence allows approximating the recommended protocol (precise TE/TR will be scanner dependent). ASSET is required to obtain correct phase, as is the disabling of any image filter and 3D geometry correction. The default 2D gradient correction works fine. Use TE = minFull. Phase Image = OFF. Must appropriately set CVs rhfiesta, rhrcctrl and rhrcxres (to the acquired matrix size).

Limitations: precise TE/TR will be scanner dependent. Export of separate channel phase and magnitude images requires a research key.

"**Product sequence**": The product SWAN and MERGE sequences allow reconstructing mag/real/imag by following the steps indicated below. MERGE allows shorter TEs and TRs. To modify the necessary CVs, a research key is required.

<u>Availability</u>: SWAN and MERGE are commercial sequences provided by the vendor.

<u>How to</u>: The sequence allows approximating the recommended protocol (precise TE/TR will be scanner dependent). ASSET (or ARC) is required to obtain correct phase, and TE must be set to "minFull". It is mandatory to set Phase Image = OFF in the GUI. The following CVs must be set as follows (research key required): rhfiesta=0 to keep echoes separated; rhrcctrl=13 to produce also the real and imaginary parts. For SWAN only, on DV25 onwards, the first echo can be imposed from the "Advanced" tab in the GUI and can be set as an integer value (CV16).

<u>Limitations</u>: No direct control on all TEs and TR.

Gradient nonlinearities on older systems (e.g. Signa HDx) lead to distortions when images are reconstructed with FFT from k-space data. These need to be corrected with the spherical harmonics of the gradient system (which are stored on the hard drive for GE).